# Title Page

## Situation Theory and Channel theory as a Unified Framework for Imperfect Information Management

Farhad Naderian; University of Applied Science and Technology; Tehran, Iran
naderian51@gmail.com; ORCID:0000-0002-1771-2325

***Abstract:***
**This article argues that the Situation theory and the Channel theory can be used as a general framework for Imperfect Information Management. There are different kinds of imperfections in information like uncertainty, imprecision, vagueness, incompleteness, inconsistency, and context-dependency that can be handled pretty well by our brain as a cognitive agent. Basic approaches like probability theory and standard logic have just an ontological point of view. So these models are intrinsically inefficient in modeling fallacious minds. The generalization of these theories to the generalized probability and nonstandard logic theories has epistemological motivations to provide better models for information integration in cognitive agents. Among many models of them, possibility theory and probabilistic logic theory are the best approaches. Most of these models are based on possible world semantics. So I argue, based on a review of different approaches to Imperfect Information Management, that a good framework for it is the Situation theory of J. Barwise and the Channel theory of J. Barwise and J. Seligman. Also, I think that Situation theory has a better epistemological foundation to refer *partiality* as one of the main sources of imperfection in information. There are many features in these theories that can be used to model and represent different kinds of information both symbolic and numeric. They have relied on a powerful and unique notion of information. So they can encompass different kinds of information with different imperfections. These frameworks have a proper approach for *context* modeling to handle *common knowledge* and *incomplete* information. Also, they distinguish *belief* from *knowledge* clearly to model the *non-monotonic* and *dynamic* nature of knowledge. They discern the logic of the world from information flow in the mind. The *objectification* process in these theories reveals to us the nature of *default* or *probabilistic* rules in perceptions. The concept of the *channel* can be used to represent those types of reasoning mechanisms that move from one model or logic to another one. The *imprecision* in our perceptions causes *fuzziness* in reasoning and *vagueness* in communication that can be represented by some suitable *classifications* connected by some *channels*. This new framework like a *network framework* can provide a *scalable* and *open* framework to cover different models based on a *relativistic* notion of truth.**

## Statements and Declarations:

This research did not receive any specific grant from funding agencies in the public, commercial, or not-for-profit sectors.

# 1. Introduction

Information has been turned into one of the fundamental concepts in this era. The need to acquire, store, process, retrieve, and transfer information effectively has motivated deep studies on the nature of the information. Amongst them, one important strand of research has emerged, dealing with the study of situations in which gathered data suffer from some imperfections, deficiencies, or uncertainties. These deficiencies in turn cause some difficulties in obtaining the required interpretations and predictions through such data. Research on the modeling of these situations is in progress by system theorists, data scientists, decision theorists, computer scientists, philosophers, psychologists, logicians, and communication engineers. Although there are many different models and approaches by the researchers to represent different kinds of information and their imperfections, there is no unified framework to cover and represent them uniquely. A unique framework is necessary to give us useful tools and methods for integrating or fusing different kinds of data. This integration or fusion is necessary not only for obtaining more reliable knowledge but also to facilitate moving from one logic or model to another.

Most theories and approaches to imperfect information handling are based on working on one old theory and then extending it with the elements of other ones. Say, extending logic by probability measures or generalizing probability theory to model ignorance. I think there is another methodology that could be called the "***network framework***". I supposed that different models of information and their extensions could be gathered together by a kind of network of models. Nodes of this network are used to represent different information models and relations between nodes are some channels for translation between models. Fortunately, the foundation of this view has been proposed before by Barwise and Seligman in their Situation and Channel theories. I have analyzed and interpreted these two theories from the point of view of imperfect information handling. Then, I have shown that different kinds of information with different kinds of imperfections could be represented by different concepts and tools of these theories.

In literature, these kinds of research have been referred to as "Uncertain Information Management". But, as I will show in 2.2 the "Uncertainty" itself is a kind of those problems in data to be studied. So, I get help from the work of Smets (2009) and select the word "Imperfect" instead of "Uncertain". This is the reason for using the word "Imperfect" in the title of my article. There could be other similar titles like "Deficient Information Management", "Imperfect Information handling" etc. Many of these titles are similar and many are not precise for our purpose. To be closer to the present literature in this research area, I take "Imperfect Information Management" to refer to those models, approaches, and frameworks that are trying to overcome processing and reasoning problems in handling imperfect information.

There are many classifications and reviews on this subject. A review of different aspects of this research with a cognitive approach has been studied by Naderian (1998). A classification of different kinds of uncertainties in AI research has been proposed by Voorbraok (1990), Hanks (1994), and Liu (2001). A review of different information retrieval models based on formal logic including uncertainty models has been presented by Abdulahad et al (2019). Another review on the similarity of problems in context exploitation, AI, and Data Fusion is in Snidaro et al (2019). A collection of papers on uncertainty management can be found in Beierle (2017). A good review including a general framework for information fusion is in Dubios et al (2016). A complete thesaurus of all concepts related to imperfection in data is in Smets (2009). I proposed my classification to see all aspects of this research area. In some cases, I have referred to the source of the theories that have been the fundamental one of their approaches in different models. This makes me have a historical point of view there, and henceforth try to find the very ideas behind the important and well-known theories.

To emphasize the importance of this research, let's refer to some application areas. Consider an advanced robot that wants to do complicated tasks. It must acquire information from its environment through its different sensors. Some information is not reliable. Many different kinds of problems are possible in the gathered information, like blurred or opaque pictures, imprecise distance measurements, false recognition of materials, etc. That robot must be able to do its tasks and must rely on this problematic information. So there must be a way for him to fuse different kinds of information with different imperfections. As another important group of applications, I can refer to expert systems. Consider an expert system for medical diagnosis. Just like an expert, it receives different kinds of information such as patient symptoms, patient history, and his/her complaints, different tests, and imaging results. It is expected from that expert system to have the ability to assess and combine the received data; because as all physicians know, some parts of such medical data could be incorrect. Therefore, that expert system must have a proper design to overcome such imperfections. Another important class of systems is those management systems that are going to make decisions based on different data from different sources and from different kinds of sources. Examples of this class could be the management systems of a big store, the management system of a big warehouse, management mechanisms of an aircraft carrier. Again, we encounter so many kinds of faulty information to be integrated and used in our decision-making. There are many sources on different applications of imperfection management for study. Some illustrated examples of different numerical methods are in Ristic et al (2020) and Li et al (2013) and Salahdine (2017). The application of situation theory in the realm of IoT (Internet of Things) could be read in Färber et al (2021). A review of 30 years of using uncertainty management in medical data is in Alizadehsani et al (2021).

Handling of imperfect information has been studied in some engineering areas, including "Data Fusion" and also "Uncertainty Management of Information" in AI. Data Fusion as a branch of engineering is mainly concerned with the problem of how one can attain more precise and comprehensive knowledge of a situation, even though various sensors, providing the necessary data, are sensitive only to a partial subset of that situation. Data fusion and integration involve six different levels of performance, including signals, pixels, symbols, sentences, knowledge bases, and databases. The signal and pixel levels are mostly considered in the field of signal processing (Abidi, 1992; Hall et al. 2004; Mitchell, 2012); however, the other levels are studied in the symbolic processing area in computer engineering. The problem of how to manipulate uncertainty in the symbols, sentences, knowledge bases, and databases is pursued in one of the branches of artificial intelligence known as "uncertainty management in information" (Voorbraok, 1990; Hanks, 1994; Sage, 1998; Liu, 2001; Smets, 2009; Dubios et al 2016; Halpern, 2017). As noted before, I take the title “Imperfect Information Management”.
In this article, regarding the basic psychological concept of information, a cognitive approach has been taken to proceed. Namely, we are looking for the answers to the question of how humans can come to know various situations of the world and venture on decisions and actions accordingly, despite deficient information. The symbol, sentence, knowledge base, and database levels of data fusion can be represented through logical methods; hence the set theory and the mathematical logic point of view have been chosen to observe "information", to fundamentally conceptualize the meaning of “imperfection”. The signal and pixel levels are not studied here, but we suppose that the final results of systems that process received signals or images are available as some higher-level information like some propositions plus some sort of uncertainty measure.
In section two, I go into detail about the main motivations for propounding the imperfect information problem and the reasons for the presence of deficiency, uncertainty, and other problems in information. Upon the presented reasoning, the numeric and symbolic approaches to imperfect information modeling are introduced in section 3. In section 4, I finally will sum up the results from the previous section to show the necessity of having a general framework. Then I analyze the Situation theory and Channel theory to show their capabilities in modeling different kinds of information and imperfections. I sum up the overall conclusions concerning the main question of this article in section 5 and note the necessary research for completing it.

# *2. Primary models of information and Imperfection representation*

## 2.1. Primary models of information

Dealing with the primary models of information necessitates explaining some basic assumptions as below. Suppose an intelligent being named "cognitive agent", gathering data via its sensors, perceives and understands the world as the human does. The study of information flow in such an entity or system requires knowledge of the informational structure of the world, the impact of the sensors on the perceived representation of the world, i.e. processing and perceptual errors, and the ways that information is processed and represented by the agent (memory and thought). The understanding of the world by the cognitive agent is based on these conditions. Concerning simplicity, the cognitive characteristics of human beings, as a perfect cognitive agent, are directly referred to in many cases here.
Mathematical logic and probability theory are the basic models of the world. Based on the consensus of many philosophical schools, the world is assumed to be a set of facts. Each fact consists of some objects with a relation among them. Consequently, the world can be represented as a set of propositions or facts, on which a structure is applicable. Naturally, since we wish to find the precise structure of the world facts, modeled by propositions, we take mathematical logic to model the exact structure of those facts. But, many intrinsically random phenomena in the world cannot be represented through deterministic modeling. On the other hand, there are a variety of deterministic phenomena that we are interested in just a probabilistic description of them. Therefore, "probability theory", which is involved in these cases, is also important as another model of the world.
Finally, the world can be represented by a set of propositions along with a logical or probabilistic structure. Up to this step, the problem is easy to manage and the standard logic and probability theory have manipulated the matter. If just the study of world modeling is favorable, it won't cause serious difficulties. In general, if the set of atomic propositions representing the world is shown by $A = \{F_i \,| i = 1, \dots, n\}$. Since every proposition can take true or false values, $2^n$ different states are possible. Each state is called a "possible world". A logical structure or a probability function can be defined on the set of atomic propositions or the possible worlds and then new deductions can be made to draw useful conclusions about the world.

### *2.1.1. Probability theory*

Probability theory is the most developed mathematical structure to represent uncertain reasoning. The main definition given in this context for the probability is the belief degree of an “ideal rational cognitive agent”. that is, if an ideal rational agent attributes the probability $\boldsymbol{p}$ to a proposition $\boldsymbol{A}$, then, in fact, its degree of belief to $\boldsymbol{A}$ is $\boldsymbol{p}$, with a degree of belief of $\mathbf{1}$ amounting to certainty and $\mathbf{0}$ to the certainty of the negation. The most important reason for taking this issue is the famous argument of the Dutch Book (DBA), which strongly defends the probability theory. According to this argument, “an agent’s degrees of belief should satisfy the axioms of probability” (Vineberg, 2016).
Based on DBA, not only the degree of a belief in an ideal rational cognitive agent could be represented by a probabilistic function, but also the representation should only be done in this way. This argument recognizes the probability theory as the only method of representation of the degree of belief.
The most developed model in this framework is the Bayes model. All of the uncertainty sources are considered together in this model. Information modeling and the effect of new evidence on the probabilistic state are based on Bayes' theorem. The Bayes' theorem itself is obtained from the conditional probability theorems. Conditional probability is a model for changing the probabilistic state of a set of propositions after presenting the evidence. For example, $P_E(*)$ is the result of changing the probability function $P(*)$ after accessing the evidence or information $\boldsymbol{E}$.
In section 3 I clarify the defects of the probabilistic model for our purpose.

### *2.1.2. Mathematical logic*

Applications of logical models in information modeling began through the development of artificial intelligence. According to artificial intelligence studies, every intelligent system should be based on a representational knowledge structure to store the received information suitably and perform reasoning accordingly.

When one says "a system has knowledge" it means it knows the world situation. If an individual's knowledge doesn't govern different world situations, that individual actually will not have any knowledge. But he merely will have some beliefs. Knowledge analysis based on the relations between the world and the language, as a model of the world, is in the field of mathematical logic and especially the model theory or truth theory analysis. In addition to providing a representational structure for knowledge, logic gives us the reasoning procedure of that knowledge.

Before the 1980s and due to the criticisms of the applications of mathematical logic in artificial intelligence[1], some believed that any use of logic in human reasoning models, especially in common sense modeling, was inappropriate. Consequently, they take other approaches such as connectionism to achieve more efficient reasoning methods. However, the vague structure of neural networks and weakness of different methods of reasoning on incomplete knowledge, except mathematical logic, raised further attention to the deduction-based logical approaches.

This is the reason why I have taken just a functional approach (and not the neural network approach) in this research and in selecting mathematical logic for imperfect information management. There are many difficulties to find the relation between cognitive tasks and neural counterparts. There are many types of research on this subject, but most of them have reached conflicting results. For a good review on the relation between cognitive tasks and reasoning in the brain, you can refer to Monti et al (2012). They concentrated on the role of language in human cognition and revealed that neurological studies could not prove the associations/disassociations between language and deductive reasoning. Although they finally proposed the role of language as the initial encoding of verbally presented materials; They added: "neither the mental representations formed as a result of the initial encoding nor the deductive operations themselves appear to be supported by the neural mechanisms of natural language". So they finally argued that clarifying cognitive task mechanisms based on neurological research needs more investigations. The physical/neural basis approach to cognitive research (including Imperfect information management in the brain) is in its infancy. Based on the above argument, I focus on the subject of the article based on a functional approach including some psychological reasoning.

Mathematical logic inherently includes proper ways to represent incomplete information (Moore, 1995):

$\neg p$: shows the lack of information about $\boldsymbol{p}$.
$p \vee q$: shows the lack of enough information on evaluating $\boldsymbol{p}$ or $\boldsymbol{q}$; in all logical or even Dempster-Shaffer numerical approaches.
$\forall(\boldsymbol{x})P(\boldsymbol{x})$: shows the absence of information about all $\boldsymbol{x}$. Equivalently, it is accepted until it is defeated.
$\exists(\boldsymbol{x})P(\boldsymbol{x})$: shows the lack of knowledge of which x has the property p.

Therefore, it is concluded that every system that wants to act upon incomplete information should use the above formal structures. Consequently, the representational structure of that system should utilize a suitable extension of first-order logic.

The criticisms of logical methods are explained in section 3.2. Also, it is described how new areas in logic have arisen. Let me note that a good basic reference to mathematical logic that I used in my studies is Dalen (1991).

## 2.2. Kinds of imperfection in information and their origins

Every cognitive agent has some restrictions if it wishes to understand the world through its perceptual and cognitive abilities. The agent's sensors, and their related perceptions, suffer from many problems because, besides their limited sensitivity, they cover only some small parts of the world. The cognitive agent's capability to store and process information (memory and thought) is limited. The information representation structures in the agent's mind differ from those in the real world, therefore their corresponding understandings will be different as well. In addition to these problems, the cognitive agent has to make decisions and take actions in each moment to survive, so it should rely on that imperfect information anyway. Hence, investigating the agent's model necessitates a more complete model, not only to model its external world but also the agent's perception and mind to consider the agent's cognitive limitations.

In literature, there are many works on this subject most of them introduced in the introduction. Also, Kleiter (2018) has distinguished logical, probabilistic, and statistical principles and argued that for a plausible model of human reasoning ingredients from all three domains are necessary. He also argued that all operators and inference rules infer interval probabilities in probability logic. Smets (1998) has worked on the difference between imprecision and uncertainty by referring to probability theory and possibility theory. He then gathered and classified many notions in this area in Smets (2009).

The various information imperfections are briefly explained below:

(1) ***Imprecision:*** the existence of numerous interpretations for a variable or an object.

A main problem of data is concealed in some quantitative phenomena, taking so many values or forms in the world. Such phenomena cannot be completely perceived by the sensor, Since the sensor can discern just a limited number of values or forms. Also, this could be the result of *the discrete nature of mental processes*. For instance,

[1] The main criticism was of the resolution deduction method because of the exponential increase in search space versus formula quantity

temperature, color, and magnitude are some kinds of these phenomena. This issue causes the received information to be "*imprecise*".

(2) ***Uncertainty:*** lack of perfect certainty towards the truth of a proposition.

Due to the *sensor measuring errors*, the interpretation of the agent's perceived propositions differs from the facts that caused them. So, the whole set of agent's sensor information has some errors and noises. These all give rise to "*uncertainty*" in the information received by the cognitive agent. There are two kinds of uncertainty. One derived from very indeterminism in the world. It is a kind of ontological uncertainty. The second is derived from our limitations to have the exact state of the world. It might be due to the unpredictability of a deterministic process or our intention not to grasp everything in a very complex deterministic process. This is a kind of epistemological uncertainty.

(3) ***Vagueness:*** the lingual counterpart of imprecision.

Imprecise information is considered "*vague*" in the lingual expressions the cognitive agent is using to refer to.

(4) ***Inconsistency**:* the existence of contradictory prepositions.

Due to sensor errors, different information gathered by the cognitive agent may contradict each other. The differences might be in analog values that could lead to contradictory results after computation or maybe in digital states that could be found immediately.

(5) ***Incompleteness**:* reasoning limitations to prove all true propositions derived from a set of information;

This is an essential feature of human reasoning and most reasoning mechanisms that cannot compute or prove all true results from a set of enough data. Theoretically, the cognitive agent must be able to reach new results based on that set of information. Depending on the reasoning system, this limitation might be absolute and unsolvable or might be technically infeasible. Resolving these problems needs dynamic gathering of information and upgrading processing capabilities.

(6) ***Incomplete data**:* indetermination of some propositions or some parts of propositions;

There are many cases in which we have no access to a set of required propositions for reasoning. This could be the result of our limitation to have suitable access to information sources. Sometimes, there is so much information and this prevents the cognitive agent to consider all of them. Also, in lingual information, some parts of the sentences or some of the propositions contained in the sentences may be unknown. This concept differs from *incompleteness*, which arises from reasoning limitations.

(7) ***Context dependency**:* dependency of information content to the context.

There is another important factor to be considered in the study of the mental characteristics of cognitive agents. That is the congregation and the communication of the agents, which is similar to the human congregation and communication. Each individual understands some part of the world based on his abilities and transfers this understanding to others via his lingual faculties, i.e. speaking and writing. Most of the human lingual expressions are indefinite, i.e. ambiguous and general, and for the earlier mentioned reasons, are vague and incomplete. Also in some cases, the existence of common knowledge in related contexts, makes it unnecessary to refer to that knowledge to communicate effectively. In these kinds of situations, human beings understand each other's speech. Those sentences which have different meanings or even include some unknown parts, all are interpreted in the special context where they are situated in. This *context-dependency* in human languages facilitates the full interpretation of sentences, despite lacking complete information.

(8) ***Partiality**:* inaccessibility to all sets of information or all possibilities for reasoning;

The cognitive agent encounters some other problems. One is the large amount of information existing in the world and the other is the limitations of the agent in storing information. Information is more than the cognitive agent's ability to gather and store. This causes the unavailability of some useful information for decision-making and action planning. In other words, the information is "*partial*". Partiality in information is a basic fact arising from this truth that we are in a limited and small part of the world and have no access to everything, everywhere, and every time.

Most of the theories in the realm of uncertainty management are going to give answers about how to handle and reason such different kinds of problems in information. These theories have been divided into two main categories: numeral models and symbolic models. I give a short review of them in the next section to capture the very notion of each one. I need this fundamental study to find the importance and position of each theory in my final framework.

## *3. Numerical and Symbolic approaches to Imperfect Information Management*

### 3.1. Numeric models

Numerical modeling of information imperfection started with the Bayesian method as one of the applications of probability theory by the famous Bayesian Rule of conditioning. According to this rule, an agent's prior or initial belief assigned to an assumption **A** can be updated to a final or posterior probability by the below formula:

**Definition:** Let's consider $\{E_1, \dots, E_n\}$ to be a set of all possible pieces of states that partition the sample space and **A** be an assumption:

$$P(E_i|A) = \frac{P(E_i \cap A)}{P(A)} = \frac{P(E_i).P(E_i|A)}{\sum_j P(E_j).P(E_j|A)} \tag{3.1}$$

Reasoning by the probability theory, especially in the Bayes' method, is facing great problems; as it requires a large amount of information. This model is not efficient in a restricted area of human cognition such as professional diagnosis. To understand the problem, consider an expert system for medical diagnosis in which, there are 20 various symptoms for 50 kinds of sickness. It needs $50 \times 2^{20}$ probability numbers (i.e. about 50 million) of different combinations of evidence to obtain the probability of being affected by a disease (Voorbraok 1990). A physician does not diagnose in such a manner. Using Bayes' rule in this way requires an enormous amount of memory and a high processing speed.[2]

Solving the complexity problem in the Bayesian method could be based on identifying the internal relations of different propositions of the model. Enhancing the Bayesian model is possible when some relations can be found between the various assumptions and the evidence. In one method, the causal relations between propositions are used. This has described in 3.1.1. The second method has used the rule-based structure in the design. like the Certainty Factor (CF) model[3] as in MYCIN[4]. The relations between prepositions are defined in the X→ Y forms, in which X and Y can be simple propositions, or the negation, the conjunction, or the disjunction of some propositions. Although the computational complexities in this model are less than the Bayes' model, it has some fundamental problems. One of the main criticisms of this model is the absence of a justifiable psychological explanation for it. The problem is how human information about certain situations could be included in this model.[5]

### *3.1.1. Bayesian networks*

The Bayesian network was proposed because of the computational problems in probabilistic approaches (Pearl, 2003). The underlying idea originated from the information structure and the relations between the propositions. A great reduction of the computational complexities is possible in this approach by identifying a logical structure for information to prevent the computational complexities included in probability theory. The structure is similar to a rule-based system in which a probability number is assigned to each proposition. Each node in a probabilistic network acts as a propositional variable instead of a proposition as in rule-based systems. A good feature of this model is its graphical implementation in which causal relations between random variables or conditional probabilities can be depicted by graphs. The casual network of this approach is based on a directed acyclic graph (DAG). The arcs of the graph represent the casual relations between propositions. These kinds of casual networks are also known as belief networks or probabilistic networks. Henceforth, the logical structure of the network causes simplicity in the problem description, reducing the need to store numerous conditional probability numbers as it was in the Bayes' method, and lowering the computational complexities. These features reveal that the above structure is a more powerful model for uncertainty than the Bayesian method.

The probability semantic in this model is based on probability distribution in the network as below:

**Definition:** Let's $\boldsymbol{X_i}$ is the representation of a preposition in the network and the set $\boldsymbol{pa_i}$ is the set of all parent nodes of it. Then the joint distribution of all variables in the network is given by:

$$P(x_1, \dots, x_n) = \prod_i P(x_i|pa_i) \tag{3.2}$$

In this formula, due to the limited number of parent nodes for each node, the number of parameters required grows linearly with the size of the network, but it is exponentially for the simple Bayesian method.

It must be noted that a Bayesian network is a direct representation of different propositions (the world facts or mental facts) and not the complete perception/reasoning process. There is no clear distinction between different cognitive processes in it. So it can not represent different kinds of uncertainties separately. This important feature has introduced by the famous Dempster-Shafer model.

[2] The PROSPECTOR expert system is another old instance in this area. It was designed for mine exploration and it also had an inefficient performance based on the Bayes' rule.

[3] In this model a number in the [-1,1] interval is dedicated to each proposition or rule of the rule-base, representing the degree of belief or disbelief in that proposition or rule. CF=1 shows the perfect belief; CF= -1 is the perfect disbelief and CF=0 represents a lack of information.

[4] A famous old expert system for medical diagnosis.

[5] In other words, how are the CF quantities extractable from human experiences? Or what are the CF equivalents in human experience? No answer had been given to this question. But the more fundamental problem is in the abortive results of cognitive justification of CFs, leading to some attempts to define the CFs according to the probability theory. This attempt finally reveals that all the formulas in the model are false. So, the theoretical justification of the CF model was not achieved, and other approaches have become more favorable (Abidi, 1992).

### *3.1.2. Dempster-Shaffer Model*

Another fundamental problem of Bayes' theory in particular and the classical probabilistic models in general, that is a motivating factor for some generalizations in probability theory, is their inability to consider the real and important distinction between two kinds of uncertainty, the uncertainty in the very set of probabilistic information (ontological uncertainty), and uncertainty caused by ignorance or lack of information (epistemological uncertainty). Frankly, there is not such a distinction in classical probability theory. This criticism was the most important reason for paying attention to the more appropriate numerical approaches, providing reasoning despite uncertainty.

Let's consider the probability distribution function of tossing a coin. It is known that if the coin is fair, the probability assigned to each side is 1/2. But if the coin status is unknown, then which probability distribution function will be considered? In such a situation, a uniform distribution function is used. That is, due to the lack of any argument to support a special distribution function, the same probability value is assigned to the different propositions of the sample space. This idea is called the "indifference principle" or the "maximum entropy principle". As the lack of information prevents us from assigning the correct probability distribution functions, that principle is used to find an optimum distribution function (Grandy et al, 1991; Kone, 1991).

However, there is no reason to assume that the real probability distribution function matches that function suggested by the principle. To be skeptical, it is better and rational to find out the upper and lower bounds of the probability distribution in the absence of enough information. This approach is considered in the generalized probability theory of Dempster-Shaffer. Keep in mind that the ignorance problem might be the result of sensor errors and inaccessibility to the complete information. (Cuzzolin, 2011)

An important class of such situations occurs when one attempts to obtain the probability distribution function in space $\mathbf{\Omega}$, while the available information is given in space $\mathbf{\Theta}$ by another probability distribution function. $\mathbf{\Omega}$ and $\mathbf{\Theta}$ are different, although they are related to each other in some way. A good example in this area is the problem of finding the probability distribution function of some of the world events when the reliability values of the sensors for perceiving those events are available. This model was proposed by Dempster at first. The sample space $\mathbf{\Omega}$ in this model is called the "target information level" and the sample space $\mathbf{\Theta}$ is called the "original information level". $\mathbf{P}$ is a defined probability distribution function on $\mathbf{\Theta}$ and $\mathbf{\Gamma}$ is a function from $\mathbf{\Theta}$ onto $\mathbf{2^{\Omega}}$. The question is what could be said about the probabilities of subsets of $\mathbf{\Omega}$, with the knowledge of $\mathbf{P}$ and $\mathbf{\Gamma}$. The epistemological uncertainties have been separated clearly from the ontological ones in this model, Since the probability function $\mathbf{P}$ over $\mathbf{\Theta}$ is the uncertainty model of a cognitive agent for real events, and $\mathbf{\Gamma}$ models the performance function of a cognitive agent's sensors. Having these two functions makes it possible to recognize the probability distribution function governing the objectively real world or the space $\mathbf{\Omega}$ to some determined extent. Further explanation of the model begins with the following definition:

**Definition**. Assume that $\langle \Theta, 2^{\Theta}, P \rangle$ is a probability space, $\mathbf{\Omega}$ is a sample space, and $\mathbf{\Gamma}$ is a function satisfying the condition:

$$\Gamma: \Theta \rightarrow 2^{\Omega}/\{\phi\} \tag{3.3}$$

then, the $P_{\Gamma,low}$ and $P_{\Gamma,up}$ functions are defined on $\mathbf{\Omega}$ as:

$$\begin{aligned} &For\ Every\ \ A \subseteq \Omega\,, P_{\Gamma,low}(A) = P(\{\theta \in \Theta :\ \Gamma(\theta) \subseteq A\}) \\ &For\ Every\ \ A \subseteq \Omega\,, P_{\Gamma,up}(A) = P(\{\theta \in \Theta :\ \Gamma(\theta) \cap A \neq \phi\}) \end{aligned} \tag{3.4}$$

The $\boldsymbol{P_{\Gamma,low}}$ function is a "lower probability function" or a "belief function", and the $\boldsymbol{P_{\Gamma,up}}$ function is an "upper probability function" or a "plausibility function". Thus, instead of obtaining a single known function, there are two probability functions, related to the probability distribution function **F** as below:

$$P_{\Gamma,low} \leq F(A) \leq P_{\Gamma,up} \tag{3.5}$$

The difference between $\boldsymbol{P_{\Gamma,low}}$, and $\boldsymbol{P_{\Gamma,up}}$ has resulted from the lack of information in space $\mathbf{\Omega}$; consequently, it can be a measure of ignorance of the cognitive agent.

After heavy criticisms of the Bayesian method, CF model, and Probabilistic network modeling, the Dempster-Shaffer (DS) model was taken into consideration. This theory had proposed by Shaffer as a continuation of Dempster's research on reasoning using belief and plausibility functions. This theory is also called the "Evidence Theory" or the "Belief function theory". The advantages of this theory over the probability theory are as below:

- Firstly, the representation of incomplete information is possible through the logical operator OR, or $\vee$.
- Secondly, as noted in the early paragraphs of this unit, this theory can handle and model the "uncertainty" and the "ignorance" separately. This is equal to consideration of the distinction between two main kinds of imperfections (ontological uncertainty and the epistemological one).
- Thirdly, the most attractive feature of this theory is the provision of a suitable rule for combining various pieces of evidence of events. The absence of such a rule in probability theory causes some computational problems in the Bayesian models.

Despite the criticisms of the DS combining rule, this theory is recognized as an effective substitute for the probability theory by many AI researchers. To emphasize the uniqueness of the theory, the descriptions of some important characteristics of the DS theory have been introduced as in Yager (2008). In DS theory, the frame or the frame of discernment concept is used instead of the sample space concept, which is usual in probability theory.

**Definition**. A mass function or the basic probability assignment (BPA) in the frame $\boldsymbol{\Omega}$ is a function $\boldsymbol{m}: \mathbf{2^{\Omega}} \rightarrow [\mathbf{0}, \mathbf{1}]$, which satisfies the two following conditions:

$$\boldsymbol{m}(\emptyset) = \mathbf{0} \qquad \sum_{A \subseteq \Omega} \boldsymbol{m}(\boldsymbol{A}) = \mathbf{1} \tag{3.6}$$

The above definition is interpreted as follows. The mass function is a representation of a rational agent's belief in an observation. The first condition means that this agent does not assign any belief to impossible events. The second condition means that the sum of all the agent's beliefs in all events equals one (the limited observation). Based on this belief assignment to some subsets of $\boldsymbol{\Omega}$, it can be possible to define the belief degree in other subsets of $\boldsymbol{\Omega}$. The belief function is introduced as a belief agent's representative in an event; interpreted based on the mass function concept.

**Definition**: Assume $\boldsymbol{m}$ is a mass function defined on the frame $\boldsymbol{\Omega}$. The belief function $\boldsymbol{Bel}(*)$ generated by **m** is defined as:

$$For\ every\ \boldsymbol{A} \subseteq \boldsymbol{\Omega}, Bel(\boldsymbol{A}) = \sum_{B \subseteq A} \boldsymbol{m}(\boldsymbol{B}) \tag{3.7}$$

$\boldsymbol{Bel}(*)$ is the *belief degree* that can certainly be assigned to event **A**. Also, it is possible to assign other belief degrees to **A**, such as the beliefs associated with the consistent propositions to **A**. *Plausibility function* is defined accordingly to take these quantities.

**Definition**: Assume m is a mass function defined on frame Ω. The plausibility function Pl(•) generated by m is defined as:

$$For\ every\ \boldsymbol{A} \subseteq \boldsymbol{\Omega}, Pl(\boldsymbol{A}) = \sum_{A \cap B \neq \emptyset} \boldsymbol{m}(\boldsymbol{B}) \tag{3.8}$$

According to the DS theory, all the information about **A** is described by the interval [Bel(**A**), Pl(**A**)]. The quantity of Ign(**A**) = Pl(**A**)-Bel(**A**) shows the degree of ignorance about **A**. The two functions Bel(**A**) and Pl(**A**) represent the maximum and the minimum probability quantities assigned to **A**, respectively. The probability function **P** is between Bel(**A**) and Pl(**A**). If complete information is available, the Ign(**A**) is zero. In this case, the belief function equals the probability function. In general, Bel(**A**) and Pl(**A**) functions are dual of each other.
An interesting feature of the approach is its provision of the evidence combination rule. Suppose there are two different pieces of evidence for a single event, represented by $\boldsymbol{m_1}$ and $\boldsymbol{m_2}$. The question is how the related belief function which resulted from the combination of these two pieces of evidence can be computed.

**Definition**: If $\boldsymbol{m_1}$ and $\boldsymbol{m_2}$ are two mass functions on $\boldsymbol{\Omega}$ and

$$\boldsymbol{E} = \sum_{X \cap Y = \emptyset} \boldsymbol{m_1}(\boldsymbol{X}).\boldsymbol{m_2}(\boldsymbol{Y}) \qquad \boldsymbol{N} = \sum_{X \cap Y \neq \emptyset} \boldsymbol{m_1}(\boldsymbol{X}).\boldsymbol{m_2}(\boldsymbol{Y}) \tag{3.9}$$

then the function **m**, representing the combination of $\boldsymbol{m_1}$ and $\boldsymbol{m_2}$, is defined by:

$$\boldsymbol{m}(\emptyset) = 0 \qquad \boldsymbol{m}(\boldsymbol{A}) = \frac{1}{\boldsymbol{N}} \sum_{X \cap Y = A} \boldsymbol{m_1}(\boldsymbol{X}).\boldsymbol{m_2}(\boldsymbol{Y}) \tag{3.10}$$

The result is represented by $\boldsymbol{m_1} \oplus \boldsymbol{m_2}$ and called the orthogonal sum of $\boldsymbol{m_1}$ and $\boldsymbol{m_2}$. The denominator of the fraction is called the normalization constant. If this number is zero, then the orthogonal sum is not defined. In other words, $\boldsymbol{m_1}$ is the perfect contradiction of $\boldsymbol{m_2}$. The degree of the contradiction between $\boldsymbol{m_1}$ and $\boldsymbol{m_2}$ is defined by:

$$wett(\boldsymbol{m_1}, \boldsymbol{m_2}) = \log\left({}^{1}/_{\boldsymbol{N}}\right) = -\log(1 - \boldsymbol{E}) \tag{3.11}$$

The more the contradiction between $\boldsymbol{m_1}$ and $\boldsymbol{m_2}$, the greater the $wett(\boldsymbol{m_1}, \boldsymbol{m_2})$. Two of the important and interesting properties of the orthogonal sum operator are the commutative and associative properties.
The main condition for using the Dempster combination rule is the independence of evidence. In data fusion, this means independence of sensor performances. Another necessary condition is the appropriate decomposition of the space of evidence, performed by the related frame function such as the evidence relationships be determined. One of the problems in DS is that the independence of the evidence set, introduced as the necessary condition for using the rule, is not defined appropriately.
Up to this point, I have reviewed the probabilistic models. The basic probability model has no clear distinction between probability in the real events and the probability number that has been assigned to cognitive agent belief about real events. This problem has been solved by DS theory in which the real-world event model has separated from the belief function. Both of these approaches have computational problems. In addition to these mentioned criticisms of probabilistic approaches, another problem with them for modeling human information integration is the non-adaptability of probability theory expressions and reasoning with lingual expressions and reasoning, respectively. Psychological studies have shown that probabilistic human judgments are inconsistent with the probability rules (Wagman, 2003). These problems in probabilistic-based approaches caused an increased interest in the approaches that are based on fuzzy logic. It seems that a better interpretation of uncertain reasoning has been presented in the area of the possibility theory (Zadeh, 1987).

### 3.1.3. Rough Sets Theory

One of the tools for handling imperfect information is the Rough set theory of Pawlak (2002). A review of this theory can be found in Skowron, et al (2018). Rough set theory mainly handles imprecise information. It is an important approach in Artificial intelligence, Cognitive science, data mining, and many other application areas. The main advantage of rough set theory in data analysis is that it does not need any additional information about data like probability numbers in statistics, probability number assignments needed in evidence theory, a grade of membership, or the value of possibility needed in fuzzy set theory.
To formally introduce this model, let's first define the information system as below:

**Definition:** An information system is a pair $\boldsymbol{S} =< \boldsymbol{U}, \boldsymbol{A} >$, where **U** and **A** are non-empty, finite sets named the *universe*, and the set of *attributes* respectively. With every attribute $\boldsymbol{a} \in \boldsymbol{A}$ we associate a set $\boldsymbol{V_a}$ of its *values*, called the *domain* of $\boldsymbol{a}$.

Any information system can be imagined as a data table with rows named by objects and columns named by attributes. Any pair $(\boldsymbol{x}, \boldsymbol{a})$, where $\boldsymbol{x} \in \boldsymbol{U}$ and at $\boldsymbol{a} \in \boldsymbol{A}$ is a particular entry in the table indicated by the value $\boldsymbol{a}(\boldsymbol{x})$. After defining the information system, the set **U** can be partitioned into its elementary sets or *granules* by an important equivalence relation defined on it. This is based on categorizing the members of U according to their indiscernibility to identify the similarity or differences in the information.

**Definition:** Any subset $\boldsymbol{B}$ of $\boldsymbol{A}$ determines a binary relation $\boldsymbol{I}(\boldsymbol{B})$ on **U**, called an *indiscernibility* relation, defined as follows:
$(\boldsymbol{x}, \boldsymbol{y}) \in \boldsymbol{I}(\boldsymbol{B})$ if and only if $\boldsymbol{a}(\boldsymbol{x}) = \boldsymbol{a}(\boldsymbol{y})$ for every $\boldsymbol{a} \in \boldsymbol{A}$.

Based on this equivalence relation, all members of **U** that are in *indiscernibility* relation with a member **x** are represented by $\boldsymbol{B}(\boldsymbol{x})$. Also, if $(\boldsymbol{x}, \boldsymbol{y}) \in \boldsymbol{I}(\boldsymbol{B})$, we say that **x** and **y** are ***B**-indiscernible*. Based on these definitions, we can define the most important notion in this theory to be used in representing imprecision:

**Definition:** The following operations on sets $\boldsymbol{X} \subseteq \boldsymbol{U}$, given by:

$B^*(X) = \{x \in U: B(x) \subseteq X,$ B-Lower approximation of X.
$B_*(X) = \{x \in U: B(x) \cap X \neq \emptyset,$ B-Upper approximation of X. (3.12)
$BN_B(X) = B^*(X) - B_*(X)$ B-Boundary region of X.

Therefore, with every rough set, we associate two crisp sets, called lower and upper approximation. Intuitively, the lower approximation of a set includes all elements that surely belong to the set, and the upper approximation of the set constitutes all elements that possibly belong to the set. The boundary region of the set consists of all elements that cannot be classified uniquely as belonging to the set or as belonging to its complement, concerning the available knowledge. If the boundary region of **X** is the empty set, then the set **X** is considered crisp concerning **B**; otherwise is referred to as rough.
According to this model, it is possible to model *imprecision* in information. Based on the above definition of rough set theory elements, it can give a suitable independent definition for imprecision. I mean that it models and represents just imprecision in data in an abstract form without any interference with other kinds of imperfection models. *Imprecision* refers to the limitation in the sensors of cognitive agents and uncertainty refers to randomness in the very phenomenon we are checking or in the error in our sensors. For my research in this article, the Rough set theory is important and interesting because it points to *imprecision* by separating it from other forms of information imperfection.

### 3.1.4. Fuzzy Logic

Imprecision or vagueness is one of the characteristics of human perception as was mentioned earlier. Research on the presence of vagueness in natural languages is not a new matter, but it goes back thousands of years ago. Here there is an introduction to the nature of vagueness in natural languages and perceptions alongside fuzzy logic as a model of vagueness. Its efficiency challenges are argued as well. There is a brief review of fuzzy logic's contribution to modeling natural language semantics in Novak (2021). As summarized in it, L.A. Zadeh first noticed that semantics of the natural language are vague in principle and suggested fuzzy set theory to model it. He then introduced some special operators to model linguistic hedge. Further research on his idea, has extended to other natural language phenomena including nouns and verbs.
The model that propounded to involve reasoning with vague expressions is fuzzy logic or fuzzy set theory. It includes the representation of vagueness in natural languages. According to Zimmermann (2010), Fuzzy logic essentially provides a natural way of dealing with problems in which the source of imprecision is the absence of sharply defined criteria of class membership rather than the presence of random variables. Imprecision here is meant in the sense of vagueness rather than the lack of knowledge about the value of a parameter.
The start of the formal definition of fuzzy logic is the definition of the fuzzy set as below:

**Definition:** If $\mathbf{X}$ is a collection of objects denoted generically by $\mathbf{x}$, then a fuzzy set $\widetilde{\boldsymbol{A}}$ in $\mathbf{X}$ is a set of ordered pairs:

$$\widetilde{\mathbf{A}} = \{(\mathbf{x}, \boldsymbol{\mu}_{\widetilde{\mathrm{A}}}(\mathbf{x}) | \mathbf{x} \in \mathbf{X})\} \qquad \boldsymbol{\mu}_{\widetilde{\mathrm{A}}}(\mathbf{x}) \text{ is called the membership function.} \tag{3.13}$$

This method helps us in defining sets and objects not as a determined and precise concept, but in a vague manner. For example, by a convex curve as the membership function for the set of concepts **X**={*Tall*, *medium*, *short*}, we can model the vagueness in them.

The main next important step in fuzzy set theory is the definition of logical operators. At first, fuzzy operators were defined as the generalization of crisp sets as below:

**Definition:** *Intersection* (logical **and**): the membership function of the *intersection* of two fuzzy sets $\widetilde{\boldsymbol{A}}$ and $\widetilde{\boldsymbol{B}}$ is defined as:

$$\mu_{\widetilde{\mathrm{A}} \cap \widetilde{\mathrm{B}}}(\mathrm{X}) = Min\left(\mu_{\widetilde{\mathrm{A}}}(\mathrm{x}), \mu_{\widetilde{\mathrm{B}}}(\mathrm{x})\right) \quad \forall x \in X \tag{3.14}$$

**Definition:** *Union* (**exclusive or**): the membership function of the *union* is defined as:

$$\mu_{\widetilde{\mathrm{A}} \cup \widetilde{\mathrm{B}}}(\mathrm{X}) = Max\left(\mu_{\widetilde{\mathrm{A}}}(\mathrm{x}), \mu_{\widetilde{\mathrm{B}}}(\mathrm{x})\right) \quad \forall x \in X \tag{3.15}$$

**Definition:** *Complement* (**negation**): the membership function of the *complement* is defined as:

$$\mu_{\widetilde{\mathrm{A}'}}(\mathrm{X}) = 1 - \mu_{\widetilde{\mathrm{A}}}(\mathrm{X}) \quad \forall x \in X \tag{3.16}$$

Due to some theoretical problems and also the necessity of diverse applications of fuzzy logic, many other forms of these operators have been defined. Regarding the cognitive basis of fuzzy logic, there is a review in Zimmermann (2010) by introducing different fuzzy logic operators. These extended forms of the logical "**and**", and "**or**" are named **t-norm** and **t-conorm**. According to him, neither one could alone cover the scope of human aggregating behavior. It is very unlikely that a single operator can model appropriately the meaning of '**and**' or '**or**' context independently, that is, for all persons, at any time and in each context.

Some ways to remedy this have been proposed in the literature. The '**min**', '**product**', '**geometric mean**', and the **γ-operator** have been tested empirically and it has turned out that the **γ-operator** is the best and context-dependent model for the '**linguistic *and***' which lies between the '**logical *and***' and the '**logical *exclusive or***'[6]. Limited empirical tests have also been executed for shapes of membership functions, linguistic approximation, and hedges. In all forms of fuzzy reasoning, the **implication** can be modeled in many different ways. Which one is the most appropriate, can be evaluated either empirically or axiomatically. Also, the models for **implication** can be chosen concerning their computational efficiency. Finally, it does not guarantee that the proper model can be chosen for a certain context.

There are also some philosophical problems with fuzzy logic. The famous one is the *Sorites paradox*. The Sorites paradox ascribed to one of Aristotle's contemporaries called Oblidous, presented an argument leading to the falsity of a proposition, assumed to be true at first. For example, from the premises that "number 2 is a small number", and "adding two small numbers does not result in a big number", it is concluded that "number 3 is a small number", "number 4 is a small number" … and after generalization, oddly "any number is a small number". As another example, consider this argument. "A man having a lot of hair on his head is not bald", and "losing one hair does not make anyone bald". Repeating these premises concluded in this ridiculous statement "a man with one single hair on his head is not bald". Another name for the Sorites paradox is the "bald man paradox" for this reason (Reed, 1994).

A method of facing the problem is to determine a salient or threshold point for each fuzzy predicate **F**, in a way that **F** is supposed to be true for the values below that threshold and not true for the values greater than the threshold. For example, men with a height of more than 180 cm are tall, and with a height less than it is short. However, as no rational reason supports such a judgment[7], it seems appropriate to take another way. This implies a deep perception of the nature of this paradox.

The Sorites paradox arises from using some vague lingual predicates like small, big, bald, rich, tall, red, etc. It should be noted that the paradox does not work on the precise concepts like "1 meter and 80 centimeters" or "uncle". Another point is that vagueness is a distinct concept from ambiguity and generality. Words such as "can" and "desert" with two meanings are ambiguous. However, words like "chair", which include different kinds, are general and refer to a class of objects[8].

Different approaches have been taken to solve the Sorites paradox problem. Frege and Russell believed that vagueness is undesirable and is the source of incoherence in the natural language. So, in their opinion, vague concepts must be removed from the natural language. They believed that vagueness doesn't exist in the world, but instead it is the result of using the natural language. Unlike Ferge and Russell, early Wittgenstein's philosophy did not believe in the existence of vagueness in the language; because, as he reasoned, if it was so, humans would not be able to communicate so much time despite vague, incoherent, and inconsistent language. This view necessitates a solution to this paradox. Contrary to the latter point of view, vague expressions do not cause inconsistency, but they are necessary for explaining some properties of the universe.

The early Wittgenstein views on language might be explained as since neither our perceptions nor our precision measuring instruments can sense the world with infinite accuracy, our understandings of the world are not always fully accurate. This leads

[6] It is defined by the convex combination of the product (as a t-norm) and the generalized algebraic sum (as a t-conorm).

[7] why consider a man with 179 cm height as short and another one with 181 cm height as tall?

[8] Although general words are also vague but, what makes them vague is different from what makes them general.

to an intrinsic vagueness in the language and our perceptions. For instance, if a measuring instrument measures one's height as 185 cm, assuming no observation or measuring errors have occurred, but as the instrument has a limited resolution[9], the measured length is vague though. Generally, whenever a continuum is to be expressed by a finite set, the vagueness arises. Vagueness is inevitable since our perceptions and instruments have limited precision. Consequently, the reasoning structure of the human should contain a mechanism that satisfies the condition of reasoning despite vagueness.
Although this theory has succeeded in expert systems, researchers tried to find the theoretical solution for the sorites paradox. Some researchers believed that fuzzy logic did not achieve any satisfaction in resolving the Sorites paradox. They proposed to find a better solution to the paradox based on deeper investigations into the nature of the modus ponens (conditional statements) and their relations with the fuzzy logic (Reed, 1994). Despite these critics, there have been some tries to resolve the Sorites paradox, some of which have been summarized in Hajek et al (2003). They also take their method in it by adding some new predicate[10] to Peano's Arithmetic to be able to successfully represent the modus ponens sentences in the numerical form of Sorites paradox. Two other methods proposed in this regard will be introduced in section 4.6.
However, fuzzy logic perfectly succeeded in expert systems and in providing another approach toward uncertainty called the possibility theory, dealt with in the following.

### *3.1.5. Possibility theory*

The possibility theory is capable of being considered an important approach for information integration and uncertainty modeling since it includes a very comprehensive, appropriate language for data fusion models. The fundamental concept of this approach is a fuzzy set, in which elements of the universe of discourse or a universal set are various representations of a set of conditions. In numerical modeling, for instance, the universe of discourse is an interval of real numbers, or in symbolic modeling, the universal set is all the possible worlds.
The theory is based on the assumption that human beings are not able to immediately realize their situation perfectly based on the evidence or perceived information. But, they could just assign a possibility degree to different interpretations of that situation. Defining a fuzzy membership function on the set of various interpretations of the studied situations is a suitable method to represent this kind of information. Possibility theory provides the required tools for modifying this membership function. More information reception means more restrictions on the elements of the universe of discourse and more changes in the membership function. Henceforth, the most possible situation in the world could be identified (Zadeh, 1978; Benferhat et al, 1997).
Not only does possibility theory provide a representation of imprecise information via possibility distribution, but also it presents a way to model uncertain information (Dubois et al. 1989). The possibility distribution $\boldsymbol{\pi_x(\omega)}$ represents the uncertain information $\boldsymbol{x \in \Omega}$. If the information itself includes any error due to the perceptual errors, that is, if the information reliability to $\boldsymbol{x \in \Omega}$ equals an amount of $\boldsymbol{\lambda}$, then the primary distribution function should be modified as below:

$$\boldsymbol{\pi_x(\omega) = max(\mu_\Omega(\omega), 1 - \lambda)} \tag{3.17}$$

This formula is used to modify the information obtained from the sensors, including some errors. There is also a general framework to solve two important problems. The first is how to modify the possibility distribution with the aid of obtaining a propositional information set "**A**"; and the second is how to conjugate or fuse or integrate the information obtained from different sources, each of which is represented by a possibility distribution.
Updating the possibility distribution by information A is provided by the following conditional distribution:

$$\boldsymbol{\pi(\omega|A)} = \begin{cases} \mathbf{1} & \boldsymbol{if\ \pi(\omega) = \prod(A)} \quad \boldsymbol{,\omega \in A} \\ \boldsymbol{\pi(\omega)} & \boldsymbol{if\ \pi(\omega) < \prod(A)} \quad \boldsymbol{,\omega \in A} \\ \mathbf{0} & \boldsymbol{if\ \omega \notin A} \end{cases} \tag{3.18}$$

$\prod(\boldsymbol{A})$ is equal to the possibility of statement **A**.
There are two ways to perform the information conjugation. If all the sources are equally reliable, the fuzzy operator **AND** is used to make all the sources equally participate in the final result. But if their reliability differs from each other, the operator **OR** should be used to increase the role of the more reliable sources:

$$\textbf{AND}: \quad \forall\omega \in \Omega, \quad \pi_\wedge(\omega) = \overset{n}{\underset{i=1}{\bullet}} \pi_i(\omega)$$

$$\textbf{OR}: \quad \forall\omega \in \Omega, \quad \pi_\vee(\omega) = \overset{n}{\underset{i=1}{\perp}} \pi_i(\omega) \tag{3.19}$$

"•" and "⊥" operators have different interpretations. Their selection depends on the application. For instance, if the logical combination of the source information has to be done, *min*/*max* operators are used. In this case, the *min* operator finds the source assigning the least possibility degree to **ω** (the *conjunction* combination), as it includes the most amount of information, while the *max* operator determines the source with the most possibility degree to **ω** (the *disjunction* combination) since it is the more reliable source. For independent sources, the "*product*" operation is recommended. Since it resulted in a very small total

[9] The length is in between [1.85-Δ, 1.85+ Δ] which Δ shows the precision of the instrument
[10] $At(\varphi)$ for "$\varphi$ is Almost ture" and $Fe(x)$ for "x is Feasible"

possibility degree as the product of all the possibilities when all the sources are reliable to denote the impossibility of **ω**. If not, the first definition is preferable, because the *min* operator eliminates extra information, but the product operator does not.
Data fusion methods by possibility theory do not come to an end, here. The nature of the theory implies various methods and different operators in fuzzy data fusion with their specific applications. This extensiveness and generality are the characteristics of the theories developed based on the fuzzy set theory (Zimmermann, 1996).

### 3.1.6. Information fusion

As noted before, there are many numerical models for imperfect information handling through the Data Fusion approach. By considering Data Fusion as a general approach to cover all previous models, one of the recent works is by Dubios et al (2016). They proposed a general framework for information fusion based on the minimum notion of information to be able to model and analyze fusion operators in different approaches. They proposed below notion of information:

An information item **T** will be characterized by three features: ($\boldsymbol{W}$ is the set of all possible worlds)

- Its ***support*** $\boldsymbol{S(T)} \subseteq \boldsymbol{W}$, contains the set of values of **x** considered not impossible according to information **T**.
- Its ***core*** $\boldsymbol{C(T)} \subseteq \boldsymbol{W}$, contains the set of values considered fully plausible according to information **T**.
- Its ***induced plausibility ordering***: If consistent, information **T** induces a partial preorder $\gtrsim_T$ (reflexive and transitive) on possible values, expressing relative plausibility: $w \gtrsim_T w'$ means that **w** is at least as plausible as $\boldsymbol{w'}$ according to **T**. We write $\boldsymbol{w \sim_T w'}$ if $\boldsymbol{w \gtrsim_T w'}$ and $\boldsymbol{w' \gtrsim_T w}$.

Then they proposed some properties for the information fusion operator $\boldsymbol{T = f_n(T_1, \ldots\ldots, T_n)}$ including the below items:

- **Unanimity**: accepting what is unanimously possible, and rejecting what is unanimously impossible.
- **Information monotonicity**: fusing a set of information, should not produce a more informative result.
- **Consistency enforcement**: fusing consistent information should give a consistent result.
- **Optimism**: in the absence of specific information, all consistent information sources are considered reliable.
- **Fairness**: all input items participate in the result.
- **Insensitivity to vacuous information**: non-informative source is useless.
- **Commutativity**: the process of fusion is symmetric.
- **Minimal commitment**: the result of the fusion should be the least informative among other possible results.

Based on these criteria, they analyzed different combination rules in different approaches to uncertainty management including set-theoretical, possibility theory, evidence theory, propositional logic, and probability theory. For example, they proved that among different combination rules in the evidence theory, below combination rule as in their previous work Dubios et al (1998) satisfy all mentioned criteria:

$$m_{DP}(A) = \begin{cases} \sum_{X,Y,X\cap Y=A} m_1(X).m_2(Y) + \sum_{X,Y,X\cup Y=A,X\cap Y=\emptyset} m_1(X).m_2(Y) & if \quad \forall A \neq \emptyset \\ 0 & if \quad A = \emptyset \end{cases} \qquad (3.20)$$

This combination rule keeps the mass $\boldsymbol{m_1(X).m_2(Y)}$ on both $\boldsymbol{X \cap Y}$ or $\boldsymbol{X \cup Y}$ if they are not empty.
The Dubios et al (2016) unification approach clarifies many capabilities and problems of all numerical approaches and some logical ones. This approach is founded on possible worlds semantics and as I will argue in section 4.1, situation-based approaches are those that I prefer for a unique framework for imperfect information management.

## 3.2. Logical models

Human beings can reason about the world based on their incomplete, noisy information at an outstanding speed. The achieved conclusions are astonishingly reliable and useful, although they resulted from incorrect and even inconsistent beliefs. It has been passed for more than three decades that artificial intelligence has realized the importance of studying the reasoning mechanism in these conditions. Non-monotonic reasoning (NMR) as an appropriate approach for modeling our intuition in reasoning with incomplete and inconsistent information has been propounded. Propositional logic and First-order logic (FOL) are not suitable structures for modeling human reasoning with imperfect information. In this chapter, studying the nature of practical knowledge accompanied with the signing of FOL inefficiency, the necessity of applying new logical models is described. First, the fundamental and important structures of the NMR are introduced based on the works of Brewka (1992).
As mentioned before, one of our practical knowledge features is its incompleteness, leading to the inability to deduce new conclusions based on it. However, we have to make decisions and take action despite the lack of information. The standard logic is not provided to be used in reasoning in this condition. The reason is the difference between monotonic and non-monotonic reasoning. If cons ( Γ ) indicates the deductible formula set of Γ:

$$cons(\Gamma) = \{\varphi \mid \Gamma \vdash \varphi\} \qquad (3.21)$$

Then monotonicity and non-monotonicity are defined as follows:

$$\begin{array}{llll} \text{If} & \Gamma \subseteq \Gamma' & \text{then} \ cons(\Gamma) \subseteq cons(\Gamma') & \text{Monotonic reasoning} \\ \text{If} & \Gamma \subseteq \Gamma' & \text{then} \ cons(\Gamma) \not\subset cons(\Gamma') & \text{Non-monotonic reasoning} \end{array} \qquad (3.22)$$

All of the types of standard logic, and among them FOL, are monotonic. These logical structures are mainly applied in describing the firm and perpetuated mathematical realities. On the contrary, the truth of practical information and world events are not firm and perpetuated, but the world is always changeable and human, in turn, gradually discovers the world. It is very common to see that new evidence discoveries cancel previous assumptions and conclusions. While it is not the case in mathematics, adding new evidence does not result in the rejection of the previous conclusions in monotonic reasoning. The general structure of monotonic reasoning is idiomatically called "permissive". That is, "if $Q_1, Q_2, \ldots, Q_n$ are theorems, then P is a theorem":

$$\frac{Q_1, Q_2, \ldots, Q_n}{P} \tag{3.23}$$

In this structure, adding new assumptions or achieving new information, does not reject the previous conclusions. Whereas the human reasoning structure is not permissive, it is restrictive; i.e. the present conclusions may be canceled by appearing new evidence. In other words, the new evidence mostly has a limiting role. The rules associated with this kind of reasoning are generally described as "if $Q_1, Q_2, \ldots, Q_n$ are theorems, and $Q'_1, Q'_2, \ldots, Q'_n$ are not theorems, then P is a theorem":

$$\frac{Q_1, Q_2, \ldots, Q_n ; \text{not}(Q'_1), \text{not}(Q'_2), \ldots, \text{not}(Q'_n)}{P} \tag{3.24}$$

Another feature of our practical knowledge is information inconsistency, the situation that resulted from gathering information from falsified, noisy sources. We, as human beings, simply solve the inconsistent problem and perform the reasoning. But what is the logical basis structure for it?
It is not allowed to use classical logic for reasoning with inconsistent information either. Since it is possible to deduct any formula from a set of FOL formulas if there is inconsistency in them. This is not the case in actual human reasoning:

$$T \text{ is inconsistent } \Rightarrow \forall \varphi \, (\Gamma \vdash \varphi) \tag{3.25}$$

On the other hand, reasoning with inconsistent information is also non-monotonic. As an instance, assume that our belief set includes inconsistent propositions { p,¬p, q, r, s }. Naturally, inconsistency in this set does not cause to set them all apart. On the contrary, we can obtain consistent subsets and deduce upon them. The above assumption set includes two maximal consistent subsets as; {p, q , r , s } , { ¬ p , q , r , s }. q, r, s, propositions are deducible from both these two theories. So, one can make conclusions such as q and r/\s. Assume that the new assumption of ¬q is added to the above set. Naturally, it is not possible to conclude q in this case. This means that adding new evidence to the set cancels some of the previous conclusions, and consequently, reasoning with the inconsistent information is non-monotonic.
There are three old problems in AI that can be resolved essentially by taking NMR as one of the most features of human reasoning mechanisms (Thomason, 2018):

- **Frame problem**
  *How could it be represented appropriately that most things remain unchanged while a happening is occurring?*
- **Qualification problem**
  *How can appropriately represent that actions might fail?*
- **Closed World Assumption (CWA)**
  *The assumption of completeness of the accessible positive information of deductive databases.*

Frame problem asks why we assume that most things will be unchanged by the occurrence of an action in a complex world with many features. A solution to this problem is to apply the "**persistence default principle**":

***"No event typically changes realities unless it is mentioned clearly".***

The non-monotonicity is inside the solution. The frame problem and the above solution made most of the researchers interested in non-monotonic logic areas.
The second problem is explained by an example. For instance, when you start your car, you expect your car to operate. However, considering daily experiences, we know that something might go wrong. The petrol tank might be empty; the battery might be out of charge; the starter might not work; and tens of other possibilities. The list of all the unusual cases may be too long, and no one either knows or tests the item a priori. Our action is based on the principle that:

***"everything goes as usual",***

and that is to apply default rules as a kind of NMR.
The last one is seen mainly in databases. Suppose there is no relation between Paris and Bonn in a database containing the information of airlines between cities. How could it be interpreted? Does it mean that there is an airline, but not mentioned? Or it means that there is not any airline at all. Usually the latter is taken. That is,

***"lack of information indicates negative information."***

This assumption is called "**the closed-world assumption**". If this assumption is not taken, we have to mention explicitly the negative information (such as the lack of airlines between Paris and Bonn), which leads to a bulk increment of the database. The existence of the useful assumption of CWA in AI causes the non-monotonicity of the proof process in databases that contain the assumption. Therefore, the reasoning on databases needs to be joined to NMR research.

Along with some other subjects, including the non-monotonic reasoning in "prolog with a not as a failure", and also the frame systems, show the importance of NMR in AI. They also model human reasoning processes despite lacking complete information. Investigations into this area began seriously in 1979, aimed generally at NMR formalization at the same level as FOL. Despite the various advantages of non-monotonic reasoning, there are many complexities involved in the formalization of non-monotonic reasoning, some of them are enumerated herein.

The recursive definition of proof is one of the characteristics of classical logic. Assuming the principle set Γ, the proof of T can be shown by a sequence of propositions. The final element of this sequence is proposition T. Each element is deduced from previous elements using the deduction rules. The primary elements are principles. Hereby, the obtained theory deduced from Γ can be determined by recursive definitions. Therefore, some algorithms are achieved for drawing out the provable formulas.

However, non-monotonic reasoning doesn't have a simple structure. The essence of reasoning in it is non-recursive. Hence, the non-monotonic theories are indicated by "fixed point" definitions; that is, the produced theory T from the principals Γ is defined as follows:

$$T = f(T, \Gamma) \tag{3.26}$$

In which, T is the result of applying reasoning principles to the information, including itself. The set T is observed on both sides of equality. From the practical point of view, this causes a very complicated, time-wasting process that requires a large amount of memory. Despite the stated limitations in using NMR, its advantages cause a wide range of investigations on the calculation complexities of NMR. Henceforth, research on using parallel processing methods, such as neural networks, can be effective. One example of it has been represented in (Wah, 1989).

### *3.2.1. Some old models of NMR*

Based on the above subjects, it reveals that NMR is closer to the adaptable human thought mechanism in the changeable world. There is fundamentally no case in which one can present a comprehensive description of the world. Therefore, our conclusions are strictly dependent on the lack of special information. In this logical structure, there is no need to repeat all of the deduction procedures when a new assumption is added; Since it internally includes assumptions like the incompleteness in the information and the developmental nature of the information. Some of the famous non-monotonic reasoning models are studied in the next sections. But the old ones are introduced briefly in this section.

- **ATMS**

An instance is a practical system called TMS (Truth Maintenance System), designed for consistent, non-monotonic reasoning. Each belief in this system is represented by a node. The belief could be true or not, and the approval and disapproval of each belief are stored as well. Each of the assumptions in turn has its way of being proved or rejected. Hence, any changes in the assumptions' truth values lead to a change in the whole structure. In TMS, changes are applied in a way that preserves the total consistency of beliefs. Although they might be gathered from different sources, the goal of the system is to fuse or integrate them. It causes the TMS or the modified ATMS type of it, to require bulk memory (Laskey et al 1990). The opposing defaults might cause complexities in TMS as well. Therefore, Laskey and Lehner have used the above structure by fusing the numeric and symbolic information, along with the DS theory. They assign a number to each of the ATMS sentences which represents proof of a belief. This number shows the reliability of the source of the related assumption. (Laskey et al 1990).

- **The theorist approach**

Another approach toward non-monotonic reasoning is the Theorists approach, which handles the natural representation of classical logic (Poole, 1987). The deficiencies are then made up by using logical semantic methods. As an instance, consider the following rule:

- *All humans are innocent until proven contradictory.*

It is not possible to represent this sentence in classical logic. It is typically a true sentence, and says "we tend to conclude one is innocent when there is no evidence referring to his guiltiness". The above rule is expressed in classical logic as below:

$$\forall x(\neg \text{guilty}(x) \rightarrow \text{innocent}(x)) \tag{3.27}$$

Anyhow, the above formula does not represent the default nature of the above sentence in any way. However, the theorist approach, considering the above formula as a theory, extends it in a way that contains the asymmetrical character of the default logic. For instance, from the following set of sentences (assumptions), those that are consistent with the "¬guilty(t)", will be selected to be added to the theory:

$$\{\neg \text{guilty}(t_1), \neg \text{guilty}(t_2), \ldots, \quad \neg \text{innocent}(t_1), \neg \text{innocent}(t_2), \ldots \tag{3.28}$$

This way, we will be able to draw rational and, of course, non-monotonic conclusions from the theory, by applying semantic considerations and theory extension.

According to the above extension, all of the people $t_1, t_2, \ldots$ are innocent since there is no reason indicating their guiltiness (represented by $\text{guilty}(t_i)$ ). Naturally, if a reason has been founded on $t_{19}$ guiltiness, then $\neg\text{innocent}(t_{19})$ is derived. Consequently, this structure is non-monotonic.

- **Logic programming approach**

Another method similar to the Theorist approach is the *Logic Programming* approach, in a way that it is extended according to a special set of assumptions to provide the non-monotonic and default-based features of reasoning. The only difference between these two approaches appears in their literature. The latter one is the programming structure of PROLOG (Reiter, 1987).

For instance, this rule " $\forall x( \neg\text{guilty}(x) \rightarrow \text{innocent}(x) )$ " can be represented in two forms in the PROLOG language:

$$\begin{array}{lll} \text{guilty}(x) & \leftarrow & \neg\text{innocent}(x) \\ \text{innocent}(x) & \leftarrow & \neg\text{guilty}(x) \end{array} \qquad (3.29)$$

From the logical viewpoint, these two conditional sentences are contrapositive and equivalent to each other. But it is not the case from a Logic Programming point of view. Suppose that each of these two sentences is separately included in two different knowledge bases. The first case implies that if there is no sentence in the knowledge base equal to "innocent(a)", then "a" is guilty. The second means that if there is no sentence equal to "guilty(a)", then "a" is innocent.

If it was aimed at having a knowledge base for determining the innocents and the criminals, both structures are useful. But using the first structure is not economical since the number of innocent people is much more than the guilty ones; Additionally, if it is not mentioned that one is innocent based on the above reasoning, he would be considered guilty. On the contrary, handling the second structure leads to the simplicity of indicating only the guilty people who are fewer in number, hence a smaller knowledge base will be achieved. The above rule is mainly a default rule that has a statistical nature. In this knowledge base, if any assumption in the form like "guilty(b)" is mentioned, the previous "innocent(b)" conclusion will be canceled. This shows the non-monotonic essence of PROLOG.

- **Circumscription approach**

The final approach, which is introduced here, is the circumscription approach of McCarthy in which, as in the theorist approach, classical logic representation is applied. Also, semantic considerations are used to minimize the extensions of some of the predicates. The non-monotonic structure is established by adding some second-order sentences to the primary theory of T. The role of these sentences is to minimize those models of T that include non-minimal predicates (McCarthy, 1986). As a simple example, consider the following theory:

$$T = \text{isblack}(A) \wedge \text{isblack}(B) \wedge \text{isblack}(C) \qquad (3.30)$$

The predicated circumscription of T is like this:

$$\Phi( A ) \wedge \Phi( B ) \wedge \Phi( C ) \wedge \forall x.( \Phi(x) \rightarrow \text{isblack}(x) ) \rightarrow \forall x.( \text{isblack}(x) \rightarrow \Phi(x) ) \qquad (3.31)$$

It is obvious that now, each extension of this theory figures only A, B, and C as black objects, and looks up for black objects only among them. No other D object will be figured as a black object unless it is directly mentioned. In this case, the above formula should be changed in a way that includes the predicate of being black for D as well. That is the non-monotonic essence of the circumscription approach.

### *3.2.1. Default logic approach*

One type of reasoning in humans is reasoning by the rules including some exceptions. These kinds of rules typically describe conditions that are true in most situations. If there is no evidence against their application (incomplete information condition), one can use them almost confidently.

This kind of reasoning is called the "default reasoning" or DR. An instance of it is presented below:

- *Birds typically fly.*
- *It is not deductible that bird X does not fly.* (3.32)
- *Then X flies.*

The first sentence in the above example is a default rule; which is not always true, but it is typically true. It has a probabilistic nature. The "penguin" and the "ostrich" are the exceptions or the contradictory samples. Applying classical logic, we need a complete list of all non-flying birds, which is impossible in practice. Even if the list is accessible, the process of approving the flying of a special bird such as a "canary" needs proof to show it is not an exception to the default rule. This procedure requires plenty of time and is distinct from our reasoning intuition in these situations. Therefore, we are guided to use the default logic (Bondarenko et al. 1997). The default logic, which facilitates incomplete information representation, is very suitable for the software structures that provide queries. The application of this logic in inquisitive languages is studied in relational databases too (Cadoli, 1997).

The default reasoning conclusions are not always true, but they are typically true. If any received information proves that a special bird does not fly, naturally the previous conclusion will be canceled. It means that the previous conclusions might be canceled, having new evidence. However, accessing a wider information set does not necessarily equal reaching more conclusions. Achieving more reliable and reputable conclusions is important, and non-monotonic reasoning points to the same fact. The default logic is an appropriate instance in this case (Reiter, 1980; Freund, 1997; Halpern, 1997).
One problem with this logic is the contradictory results when different default rules get involved in reasoning. The problem is that what would be the criterion for selecting the suitable rule in this case? Another problem of the DR is the lottery paradox. Assume that N people buy the lottery tickets in which there is only one winner. If N is a large number, it may be concluded that each person uses this default rule that no typical **C** person would be the winner. Hence, the DR structure implies that no one will win, which is certainly a false conclusion. A good solution to this problem is the application of those logical structures that assign uncertainty degree or truth probability to the logical propositions, like penalty logic (Pinkas, 1995), probabilistic logic (Halpern, 1990), and fuzzy logic (Zimmermann, 1996).
Assigning real numbers to the propositions, representing their degree of uncertainty has made the penalty logic able to confront the contradictory defaults and the lottery paradox. In probabilistic logic, as it is the generalization of the probability theory from the logical viewpoint, the sample space includes the simple propositions and their logical connections. Halpern has selected two different approaches toward assigning a probability to the propositions. In the first approach, the probabilities are assigned to the domains that are under consideration. The second approach assigns the probabilities to the possible worlds. The first approach is useful in formulas that contain statistical information about world events. The second approach is beneficial to the formulas indicating a cognitive agent belief. Halpern presented a structure in which both of these approaches are applicable. Here, as in the generalized probability theory, ontological uncertainty has been separated from the epistemological one and both are dealt with a unique structure (Halpern, 1990). The possibility of applying probabilistic logic in representing the default rules has been studied by Bacchus, et al (1996). They have provided the possibility of assigning a belief degree to a cognitive agent's mental propositions by selecting an approach named "statistical worlds". Like the Halpern approach, they have used statistical information, first-order information, and default rules. Their approach is based on assigning equal probabilities to the possible worlds.
To have an intuitive sense of the combination of logic with probability theory, let's refer to the good review of these models in Demey et al (2019) and I will give a summary of some of them here. There are many extensions to logic to combine both logical and numerical information in one theory. Most of these extensions are based on probability theory. One type of these extensions is based on maintaining a logical system and then defining a probability function on their terms instead of the truth validity function. Another new approach is based on defining some suitable probability operators on logical expressions.
The first important probabilistic extension of propositional logic is *Probabilistic Semantic.* This approach is based on considering a probability function as below on the terms of propositional logic language $\mathcal{L}$:

$$\begin{array}{ll} \text{Non-negativity:} & P(\phi) \geq 0 \;\; for\ all\ \phi \in \mathcal{L} \\ \text{Tautologies:} & if\ \vDash \phi,\ the\ P(\phi) = 1 \\ \text{Finite additivity:} & if\ \vDash \neg(\phi \wedge \psi),\ then\ P(\phi \vee \psi) = P(\phi) + P(\psi) \end{array} \tag{3.33}$$

Based on these definitions, the semantic validity is defined as below:
An argument with premises $\Gamma$ and conclusion $\phi$ is said to be *probabilistically valid*, written $\Gamma \vDash_P \phi$, if and only if for all probability functions $P: \mathcal{L} \to R$:

$$\text{if } P(\gamma) = 1\ for\ all\ \gamma \in \mathcal{L}, \text{ then also } P(\phi) = 1 \tag{3.34}$$

The soundness and completeness of such logic concerning probabilistic semantics have been proved:

$$\Gamma \vDash_P \phi \quad \text{if and only if} \quad \Gamma \vdash \phi \tag{3.35}$$

The difficulty of this approach is in using deduction rules to reach from the probability of premises to the probability of the conclusion. It has been shown that the exact probability of conclusion is not always obtained from the probability of premises, but an upper and lower bound of it is feasible with a quite high computational complexity.
Another approach to integrating probabilistic rules into the logical framework is in using probabilistic operators in deductions. In one approach, the language is extended with a unary operator □, which is to be read as 'probably'. Hence a formula such as $\Box\phi$ is to be interpreted as "$\phi$ being sufficiently more probable than its negation". In another approach a binary operator ≥ has been introduced in such a way that $\phi \geq \psi$ is to be read as "$\phi$ is at least as probable as $\psi$" or '$P(\phi) \geq P(\psi)$'. These approaches are closer to normal human reasoning than previous difficult numerical approaches because people are often willing to compare the probabilities of two statements or are willing to refer to the high probability of events instead of their exact probabilities.

### *3.2.2. Approaches based on modal logic*

Another form of non-monotonic reasoning is Autoepistemic reasoning. A conventional example in this area is:

- *I know all of my brothers.*
- *I don't have any information including that James is my brother.* (3.36)
- *So James is not my brother.*

In this case, the statements of this structure point to the information one has about his knowledge. This pointing to one's knowledge is the reason for calling that "auto-epistemic reasoning". The non-monotonicity of this kind of reasoning arises from one's knowledge dependency on the context. Suppose extra information that "James is truly my brother" is received. There are two ways to confront this matter.

One is to take this proposition as the canceller of the first proposition. Hence, the first proposition has been distracted from the reasoning.

In the second approach, it is supposed that new evidence points to the knowledge content of the agent, and that evidence just causes some changes in the content of the agent's knowledge. So the auto-epistemic proposition remains in the reasoning but with new meaning or context of interpretation to contain the new evidence.

The latter approach, which is more rational, denotes the relativity in knowledge and has a non-monotonic nature because previous conclusions (say, "James is not my brother") could be canceled based on new evidence (Smets, 1998).

It is worth mentioning that in the language of this approach, expressions like $\mathcal{L}$P are used, which are true if the contradiction i.e. ¬P is not justifiable. $\mathcal{L}$ is a modal operator which takes different meanings, such as "I believe in…", "I know that…", or "It is justifiable to…". However, the most important model in this approach is the Auto-epistemic Logic (AEL), propounded by Moore (1995). AEL is fundamentally seeking to model an ideal introspective cognitive agent. This agent can make reasoning upon its known and unbeknown. That is, it knows exactly what it knows and what does not (Self-Knowledge). Henceforth, the modal operator "$\mathcal{L}$" corresponds to "I know that".

Epistemic Logic is based on standard propositional language with a standard Boolean operator like 'not', 'and', 'or', and 'if-then'. It also uses operator $K_i\varphi$ ( Agent 'i' knows that $\varphi$ is the case) and $C_G\varphi$ ($\varphi$ is common knowledge in group G). In summary, the language terms in this logic are composed iteratively according to the below rules: (consider 'p' as an atomic proposition)

$$p \mid \neg\varphi \mid \varphi_1 \wedge \varphi_2 \mid K_i\varphi \mid C_G\varphi \tag{3.37}$$

Semantic theory in Epistemic logic is based on possible world semantics as below:

**Definition:** A *model* $\boldsymbol{M}$ is a tuple $M = \prec W, R_i, V \succ_{i\in I}$, where W is a set of possible worlds, $R_i \subseteq W \times W$ is agent i's binary accessibility relation between worlds (preferably equivalence relation), and *V* is a *valuation* assigning truth values to proposition letters at worlds. ($\boldsymbol{M}$,s) means that s is the actual world in model $\boldsymbol{M}$.

Based on this semantic, the interpretations of language terms are as below:

| | | | |
|---|---|---|---|
| $M,s \vDash p$ | iff | V(p) is true at s | |
| $M,s \vDash \neg\varphi$ | iff | $M,s \vDash \varphi$ is not the case | |
| $M,s \vDash \varphi_1 \wedge \varphi_2$ | iff | $M,s \vDash \varphi_1$ and $M,s \vDash \varphi_2$ | (3.38) |
| $M,s \vDash K_i\varphi$ | iff | for all t that are accessible from s: $M,t \vDash \varphi$ | |
| $M,s \vDash C_G\varphi$ | iff | for all t that are accessible from s by some finite steps: $M,t \vDash \varphi$ | |

There are many axiomatic systems for this semantic with different completeness conditions and different applications. The most important one is named S5 with the below rules:

| | | |
|---|---|---|
| - All valid principles of propositional logic including Modus Ponens | | |
| - $K(\varphi \rightarrow \psi) \rightarrow (K\varphi \rightarrow K\psi)$ | Modal Distribution | |
| - $\varphi \rightarrow K\varphi$ | Necessitation | |
| - $K\varphi \rightarrow \varphi$ | Veridicality | (3.39) |
| - $K\varphi \rightarrow KK\varphi$ | Positive Introspection | |
| - $\neg K\varphi \rightarrow K\neg K\varphi$ | Negative Introspection | |

This framework could help us in modeling mental representations of the cognitive agent. Two main mental states that are important in this article are *Knowledge* and *Belief.* The mental state "*Belief*" has been mentioned in many previous models (probability theory, possibility theory) to refer to the uncertain information of a cognitive agent. The second important mental state is *Knowledge* which is used when a cognitive agent is going to refer to his own certain information. Two modal logic structures for modeling these mental states are AEL and AELB.

AEL is the product of propositional logic generalization by the modal operator $\mathcal{L}$ (or Operator $K$ in S5). Its goal is merely to model human conscious knowledge. However, we sometimes need to make deductions not only about our knowledge but also about our beliefs. Przymusinski develops a structure that deals with the matter called the Autoepistemic Logic of Knowledge and Belief (AELB) (Przymusinski, 1997). Based on the AELB model, we have a non-monotonic reasoning mechanism in a cognitive agent considering his belief and knowledge. This logical structure includes the knowledge operator $\mathcal{L}$ and the belief operator $\mathcal{B}$. It provides a unique framework for many of the NMR forms. It also provides the possibility of reasoning despite the presence of inconsistency in information.

One of the main causes to propound the modal logic in artificial intelligence is to make reasoning possible in a logical system containing inconsistent information. Because in these frameworks, inconsistency will be interpreted as the situation in which some contradictory information is received and that is on the reasoning mechanisms of an agent to deduce and select a consistent subset of it as its knowledge. Both AEL and AELB models have this possibility. Lin (1998) also, proposed a new propositional logic by adding two $B$ $and$ $B^C$ operators, to respectively represent "having belief" and "having consistent belief", in a way that establishes the possibility of consistent reasoning despite inconsistency. The cognitive agent will be able to reason about its consistent beliefs with the aid of $B^C$ operator, and to reason about all of its beliefs with the aid of operator $B$.
Another main approach in modal logic models is those modal logic forms that are going to consider change or dynamicity in the knowledge or belief of the cognitive agents. Some of the main models are DEL (Dynamic Epistemic Logic), PAL (Public Announcement Logic), and AGM (belief revision). There is a good review of standard models and new models in Benthem (2010).
In DEL, another model for events is added to the semantic to model the event conditions and results:

**Definition:** An *epistemic event model* ***E*** is a tuple $E =\prec E, R_i, pre_e, e \succ_{i\in I}$, where E is a set of events, $R_i \subseteq E \times E$ is agent i's binary accessibility relation between events, and $pre_e$ is the set of preconditions for event *e* and finally the actual event *e*.

This model is going to help us in identifying the event conditions for different agents. The relation $R_i$, is used to tell us which events are possible for cognitive agent *i* in each state. $pre_e$ is the set of all preconditions of each event *e*.
Then a new model named the *product model* is defined by the Cartesian product of models **M** and **E** consisting of some product rules. This new model is used to dynamically move from some old models by some events and according to some preconditions toward new models.
There are many tries to integrate modal logics with numerical approaches like probability theory. One of the good surveys in this regard is by Demey et al (2014).
One of them is probabilistic epistemic logic which is a probabilistic extension of epistemic modal logic S5. To show the representational power of this extension, I start with some definitions. Consider a finite set I of agents, and a countably infinite set Prop of the proposition. Then:

**Definition:** A *probabilistic Kripke frame* is a tuple $F =\prec W, R_i, \mu_i \succ_{i\in I}$, where W is a non-empty finite set of states, $R_i \subseteq W \times W$ is agent i's epistemic accessibility relation (preferably equivalence relation), and $\mu_i: W \to (W \to [0,1])$ assigns to each state $w \in W$ a partial function $\mu_i(w): W \to [0,1]$, such that:

$$\sum_{v\in dom(\mu_i(w))} \mu_i(w)(v) = 1 \qquad (3.40)$$

**Definition:** A *probabilistic Kripke model* is a tuple $M = \prec F, V \succ$, where F is a probabilistic Kripke frame (with a set of states W), and $V: Prop \to \wp(W)$ is a valuation.

The function $\mu_i(w)$ represents i's probabilities or degrees of belief at state w. The language of Epistemic logic then has extended by *i-probability formulas* for representing cognitive agent probabilistic inferences like the last formula below:

$$\varphi ::= p \mid \neg\varphi \mid \varphi_1 \wedge \varphi_2 \mid K_i\varphi \mid a_1 P_i(\varphi) + \cdots + a_n P_i(\varphi) \geq b \qquad (3.41)$$

The language includes normal symbols for simple propositions and also epistemic propositions. As an example for the last term, $P_i(\varphi) - 2P_i(\psi) \geq 0$, means that 'agent I consider $\varphi$ to be at least twice probable as $\psi$'. Also, $P_a(K_b\varphi) = 1$ means that 'agent a is certain that agent b knows $\varphi$'. Then a sound and complete axiomatization of probabilistic epistemic logic is introduced according to the *probabilistic Kripke frame* and related semantics. (Demey et al, 2014). They also introduce a probabilistic Dynamic Epistemic Logic by introducing a *probabilistic update model* to represent actions and related preconditions as below:

**Definition:** A *probabilistic update model* is a tuple $E =\prec E, R_i, \Phi, pre, \mu_i \succ_{i\in I}$, where E is a non-empty finite set of events, $R_i \subseteq E \times E$ is agents i's epistemic accessibility relation, $\Phi \subseteq \mathcal{L}^{\otimes}$ is a finite set of pairwise inconsistent sentences called preconditions, $\mu_i: E \to (E \to [0,1])$ assigns to each event $e \in E$ a probability function $\mu_i(e)$ over E, and $pre: \Phi \to (E \to [0,1])$ assigns to each precondition $\varphi \in \Phi$ a probability function $pre(\varphi)$ over E.

By extending the language by the above model, we have the necessary representation terms to refer to the dynamics of the states. By *probabilistic Kripke model* $M$, we refer to static information, and by *probabilistic update model* $E$, we refer to those types of events that are taking place. Then new agent's static information, in both its epistemic and its probabilistic aspects is represented by a product of $M \otimes E$. For example, the meaning of $pre(\varphi)(e) = b$ is that 'if $\varphi$ holds, then event e occurs with probability **b**'. The soundness and completeness of the related semantic and proof system have been proved. The updated formula for this model is as below:

$$\mu'_i(w,e)(w',e') \coloneqq \frac{\mu_i(w)(w').pre(w')(e').\mu_i(e)(e')}{\sum_{w"\in W/e"\in E}\mu_i(w)(w").pre(w")(e").\mu_i(e)(e")} \qquad (3.42)$$

This formula says that "agent **i** calculates at state $(w,e)$ her new probability for $(w',e')$ by taking the arithmetical product of its *prior probability* for $w'$ at w, the *occurrence probability* of $e'$ in $w'$, and also its *observation probability* for $\boldsymbol{e'}$ at $\boldsymbol{e}$, and then normalizing this product."

This approach is a new development in imperfect information management and needs more investigations for some open problems. One open problem is the applicability of plausibility function instead of probability to be integrated into the Epistemic Logic. Another important issue with this approach is the exact relation between the quantitative (probabilistic) and qualitative perspectives on soft information. It means the probability must be assigned to belief, and the probability must be assigned to knowledge.

## 3.3. Context modeling

As mentioned before, one of the sources of imperfection in information in the language and human communication is context dependency. For example, the sentence "Bob came" is meaningful in some contexts, but could not be interpreted in some other contexts due to some incomplete information. We mainly interpret information based on a set of facts in the present situation or based on previous information. If new information has no relation to previous information or has no link to the present situation, then it cannot be interpreted easily. Each piece of information can be interpreted based on some corpus of information or in a network of interrelated information. The set of information in our minds that facilitates new information interpretation is called *context*. Modeling and representing context as it is in our mind is one of the main missions in AI that started with the works of McCarthy (1990).

### *3.3.1. Situation calculus*

The first serious movement in context modeling was that made by McCarthy (1990). He had used the notion of contexts in most of his previous research, like Situation Calculus and Circumscription Theory. But, he considered context as a unique concept by defining the predicate $ist(c, p)$ which has the meaning that "proposition $\boldsymbol{p}$ is true in context $\boldsymbol{c}$".
McCarthy has not proposed a unique and formal notion of context, but he proposed to examine each context on his own. Although he believed that some contexts are too rich to be shown by situation calculus (McCarthy et al. 1998).
The basic elements of this calculus are: (McCarthy, 1963)

- The *Actions* that can be done in the world
- The *Fluents* that describe the state of the world
- The *Situations* as a finite sequence of Actions. The *Situation* consists of the history of Actions.

There are also some logical rules to show preconditions of actions, to describe situations, to derive Fluents from actions, and some fundamental axioms of situation calculus itself.
Those Fluents that have derived from a situation are like a kind of information that is dependent on that specific situation and they might not be derived from some other situations. This way, a rich context could give us more information than those situations or contexts that are different or maybe are poor.
Another useful notion in McCarty's theory to model those problems in information due to context-dependency is the *lifting axioms* (McCarthy et al. 1998). These axioms are used to relate truth in one context to truth in another context. He considered lifting axioms as informal notions that could be investigated in each case or context separately. A cognitive agent that acts based on situation calculus and has the necessary kind of lifting axioms could interpret incomplete information based on the history of actions that are in that situation. The context in this meaning is used as the source of common knowledge for that agent.
A more formal version of McCarthy's approach is in (Buvac, et al; 1993). They extend a sound and complete propositional logic by a predicate for context reasoning as $ist(\kappa, \phi_0)$ to show that $\phi_0$ is true in context $\kappa$. He also continues extending language to a first-order language enabling us to refer to properties of contexts in first-order formulas by quantification on context. (Buvac, 1996).
Another application of situation calculus that is notable here is the work by Belle (2020). He surveyed recent results about the integration of logic, probability, and actions in the situation calculus as one of the oldest and most well-known formalisms. He also explored reduction theorems and programming interfaces for the language motivated by cognitive robotics. He makes use of two distinguished binary Fluents 'p' and 'l'. The p fluent determines a probability distribution on situations, by associating situations with weights. More precisely, the term $p(s', s)$ denotes the relative weight accorded to situation $s'$ when the agent happens to be in situation s. The term $l(a, s)$ is intended to denote the likelihood of action 'a' in situation 's' to capture noisy sensors and effectors. This way he tried to enrich the language of situation calculus for logical-probabilistic reasoning in a robot.

### *3.3.2. Semantic Approaches and Situation Theory*

The first theoretical set of approaches is those that are based on truth theory to show the meaning of sentences or propositions. A brief of it is in Kempson (1998). The first model of context was offered by Montague in Intensional Logic. He proposed this logic to solve the ***opaque context*** problems in Davidson's theory. Both Montague's theory and Davidson's theory are based on truth theory.
To Davidson, the meaning of a sentence can be obtained through the meaning of its words and the meaning of words are real extensions of them that are real entities in the world. For example, the meaning of the first sentence below can be obtained by the meanings of its constituent words:

- David has an AI book in his hand.
- David is searching for the AI book.

The meaning of the second sentence cannot be obtained through Davidson's framework, because there is no extension for "AI book" in it. In fact, there may be no "AI book" at all in that context. Such contexts are named "***Opaque contexts***". There are many other examples of such contexts in semantics.

Montague proposed the notion of *Intension* to solve this problem. *The intension* is against the *extension*. Extension of a word refers to the class of real entities that are referred to by that word, but *intension* is the property of the class that is referred to by that word. For example, the *extension* of 'apple' is a member of the class of all apples of the world, but its *intension* is the property of all 'apples'. As another example, there is no *extension* for the word 'phoenix' but the property of being 'phoenix' is present, so it has *intension*.

Montague then added the world-time *index* and *possible worlds* to Davidson's theory to achieve a mathematical model of *intension* and *extension*. Each index refers to a point or area in space-time. *The extension* of a word is an object or property in the world and is defined in a specific *index*, but its *intension* is defined by the function of all *index-extension* pairs. Finally, the *possible world* is a function of each *index* to the logical status of the world in that *index*.

Based on this theory, the second sentence above has meaning; because although it may not have an *extension* for 'AI book' in the context, there could be an *intension* for it. (There is at least one index in which an *extension* of 'AI book' exists.)

The notion of context can be defined by this theory too. The context of a sentence is the set of all possible worlds that contain all information related to that sentence through considering all *extensions* and *intensions*. This context is used to interpret the newly received information, even though it contains some deficiencies just like the absence of the "AI book" in the example.

Although Montague Intensional logic has succeeded in solving the *opaque contexts* problem, it cannot handle those complicated scenarios in which one sentence is referring to one of its parts. This has been solved by Situation semantics.

The most powerful and formal notion of context modeling is in Jon Barwise's theory of Situations. Situation theory is a mathematical theory of information (Devlin, 1991). This theory, with a powerful basis on philosophical work on the nature of information and with a phenomenological approach, considers the most necessary and real structures required to model context and related problems like communication and knowledge (Jon Barwise et al. (1983). Despite the initial intention of this theory, it has been used as a mathematical theory of information (Barwise, 1989) and also as a theory for information flow in distributed systems (Barwise et al. 1998).

This theory has a powerful ontology, which permits it to model information flow and cognition, not only in a single cognitive agent but also in a system consisting of a community of agents. This modeling makes it easy to consider social aspects of information and finally, the context-dependency of information as I pursue in this article. (Barwise et al. 1987)

To justify the claim of this article, I give a useful brief of situation theory in the next section. For a better understanding of situation theory, it is useful to review some basic concepts. At each step, I reason about the usefulness of that concept for imperfect information management. A good reference to cover all necessary basic definitions of this theory is in the work by Seligman (2014).

## 4. Situation Theory and Channel Theory as a General Framework

According to the results of the previous sections, it is revealed that none of the presented approaches and models can capture all aspects of imperfect information management as described in section 2. Although there are many tries to combine some models to present a more comprehensive one as in Figure 1, they have not obtained complete success. Most numerical methods have no plausible interpretations about the certainty measure they have taken. None of them has a unique theory of information to model both numeric and symbolic information. Almost most approaches cannot model context-dependency in a good and reasonable manner. Most models are based on possible world semantics and could not handle partiality in information as is essential in human reasoning.

I am not going to criticize these models, but I am going to insist on considering all of them in a powerful and comprehensive framework which I have referred to it as a "***network framework***". In this section, I am going to introduce the Situation theory and its extended version named "Channel theory". I argue that these theories can make a comprehensive framework to encompass all different models and approaches. Also, I show that upon taking this methodology, the previously mentioned models and approaches could find their role and position in the *network framework*. I do not consider Situation theory/Channel theory as a competitor theory to previous ones, but I recognize it as a general framework with a high level of abstraction to give a common language for reasoning in a network of different models connected by channels.

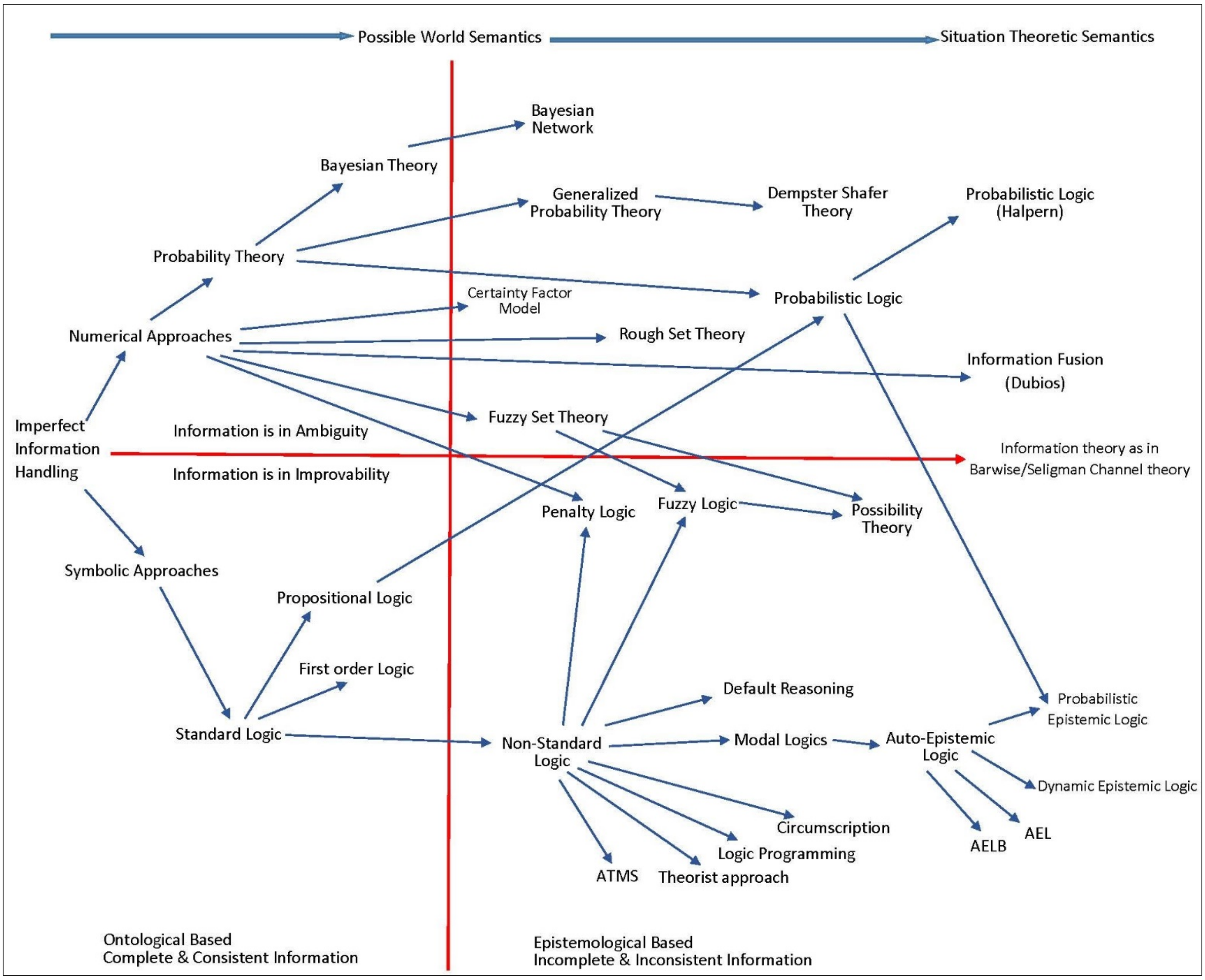

*Figure 1*: An overall overview of different models and approaches in Imperfect Information Management

As another reason to do this let's note the fact that humans are different and have different cognitive abilities. Considering one limited model to describe all cognitive abilities is not reasonable. Instead, it is better to have a general framework to encompass all models to be able to model different forms of cognitive capabilities in different theories. Different theories arise from different minds. Managers, scientists, and data analyzers can handle probabilistic knowledge easily. Mathematicians, logicians, philosophers, and lawyers engaged in logical knowledge. Normal people mainly use default rules and fuzzy deductions in their reasoning. So, there is not a unique model for all kinds of information representations and imperfections in information. Even in a single human, there are different kinds of reasoning in different situations for encountering different problems. So instead of taking one model, it is better to take a general framework to involve all kinds of information and their relations.

Also, I must note that Situation theory and Channel theory have a powerful philosophical background in modeling information, knowledge, reasoning in cognitive agents, and also imperfect information. A standard model of imperfect information management must be able to handle all kinds of knowledge and information, and it must have a mechanism to convert or translate different representations. Situation theory is the necessary theory for modeling different representations of information and knowledge and Channel theory has the suitable tools for modeling the relations between those different representations.
There is another duality in this area that are in Figure 1 and I should describe it. There are two different concepts of information that are in symbolic and numerical models or correspondingly logical and probabilistic models. In mathematical logic, we can find information in those sentences that cannot be derived from present and in-hand information, and in probability theory, information is in those events that are improbable and unanticipated for us or, technically speaking, “could not be stochastically predicted by a present probability distribution function”. I think every effort to reach a unique framework for imperfect information management should have a unique definition of information. As in Figure 1, there is a lot of research on how to model both symbolic and numeric information in one framework, but I have not found in any previous research a unique and comprehensive mathematical definition of information, which I think is necessary.
This methodology in handling the subject leads us to an ***open framework*** for accepting different theories and merging them into that universal framework. For a better explanation, remember that we (I mean human beings) are in the middle of our epistemological development and our knowledge and information about the world are developing gradually. So there is no need to have a universal and complete model of everything, but it is better to have an open model to encompass all kinds of knowledge and information and also could be augmented and developed according to new models and new theories.
A scheme of what I have told is in Figure 2. Two cognitive agents are getting information from some situations S1-S6. The related channels for this information gathering are C1-C9. There are some internal situations in the brain of cognitive agents, $S'1 - S'10$. Information about external situations is conveyed to these internal situations by related channels. Some of the internal situations are not connected to external situations but are connected to some other internal situations. Channels C8-C9 are special because they constitute communication channel between two agents. S4 is a shared situation of two agents. S5 is another shared situation of two agents for their lingual communication. In channel theory, the situations that are connected by the channel are named classification as the definition in 4.4.
Another important feature of the ***network framework*** proposed here is ***scalability***, just like any other networking plan. Scalability is one of the important features of an open framework and it let us develop and evolve the system without any severe changes in the previous structure. By taking this plan for imperfect information handling, gradual evolution of the model will be achieved and this is very important for any general framework to be accepted by experts.

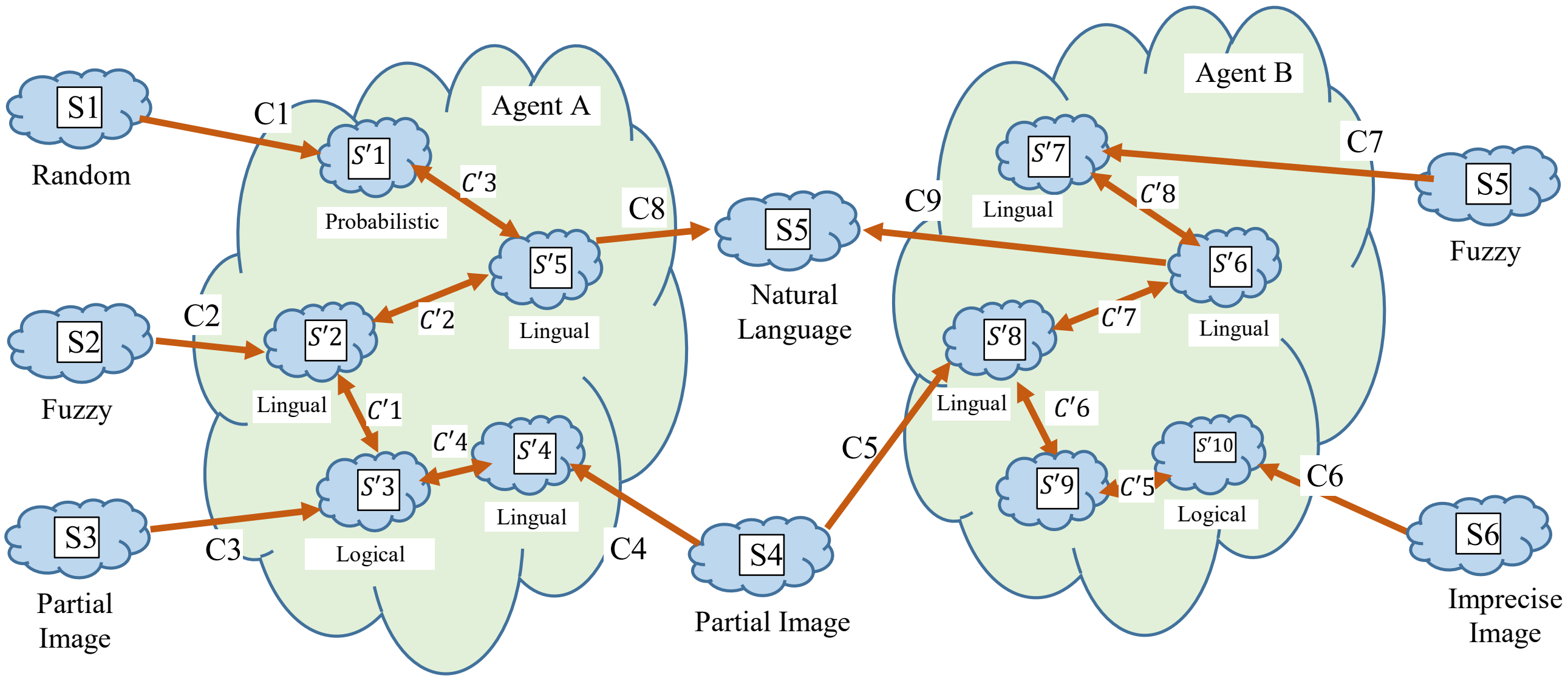

***Figure 2***: **a *Scalable* scheme of *Network Approach* based on Situation Theory and Channel theory**

Accepting this framework, we could get rid of so many difficult modeling of combined approaches. Then a wide and astonishing research area will be started. These new researches will concentrate on modeling different channels or translations between different information models. Developing the network framework is not easy at all, but I try to give some more justification on it based on its motivating theories, Situation theory, and Channel theory. Below I mention a list of different theories as in sections 2 and 3 together with their counterparts in the new framework:

| Model/Approach | Imperfection Modelling | Network Framework |
|---|---|---|
| Probability Theory | Uncertain situations in real-world<br>Belief functions in the mind<br>Perception error model | 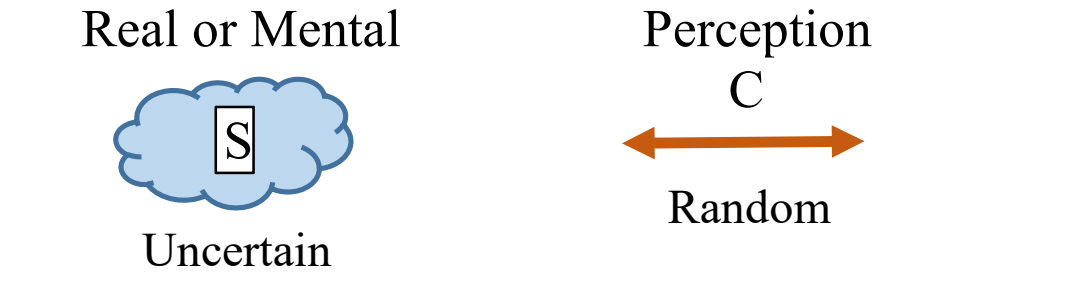 |
| Bayesian Theory | Uncertain situations in real-world<br>Belief functions in the mind<br>Channel has no error | 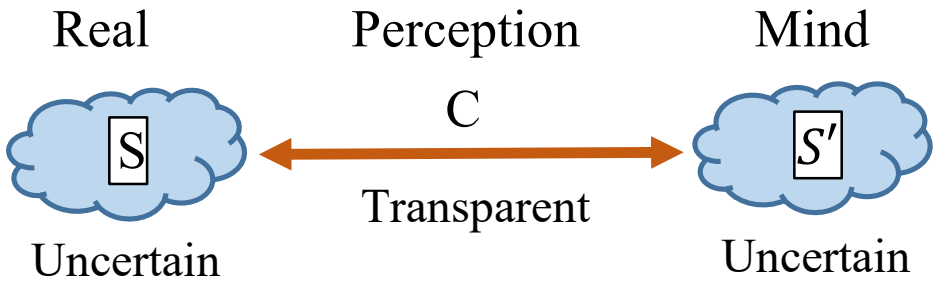 |
| Dempster-Shafer Theory | Uncertain situations in real-world<br>Belief functions in the mind<br>Channel has no error<br>Combination rules to consider some sources<br>Channels can be uncertain | 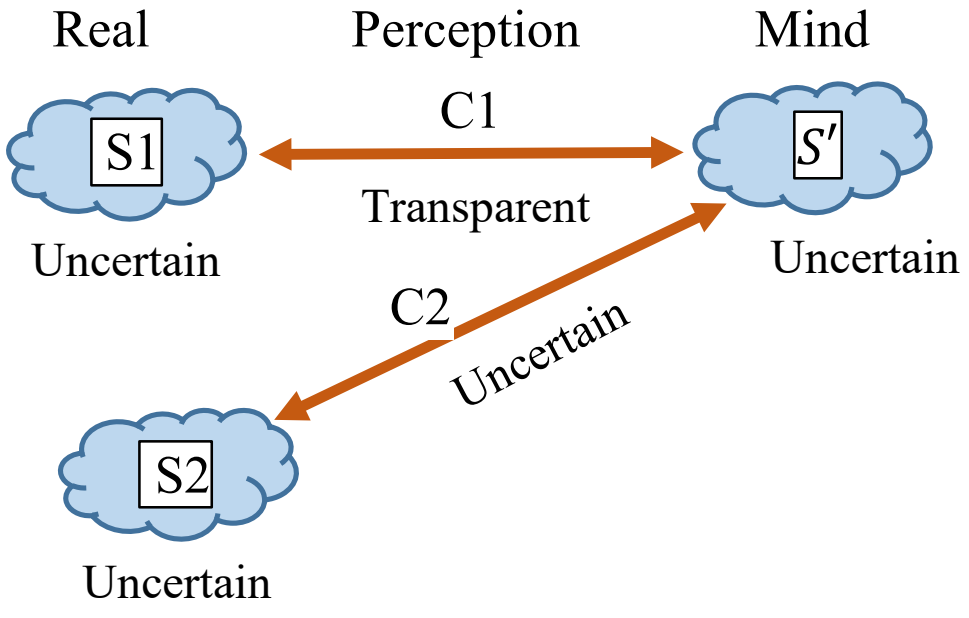 |
| Rough Set Theory | Imprecise data set in real-world<br>Imprecise Channels | 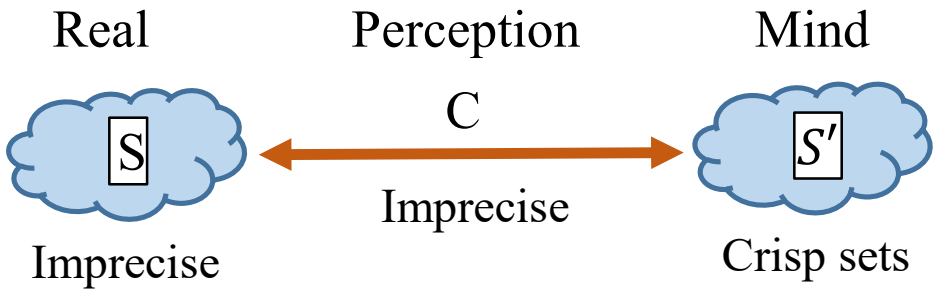 |
| Fuzzy Set Theory | Fuzzy data set in real-world<br>Fuzzy Channels<br>Vague/Crisp model in the Mind<br>Probabilistic channels | 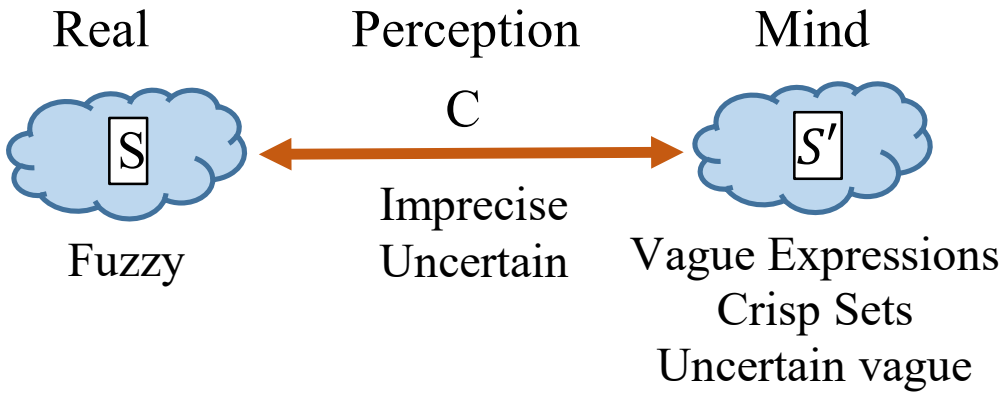 |
| Possibility Theory | Imprecise/Fuzzy data set in real-world<br>Imprecise/Fuzzy/Uncertain Channels<br>Vague/Crisp/Uncertain model in the Mind<br>Combination rules to consider some sources<br>Probabilistic channels | 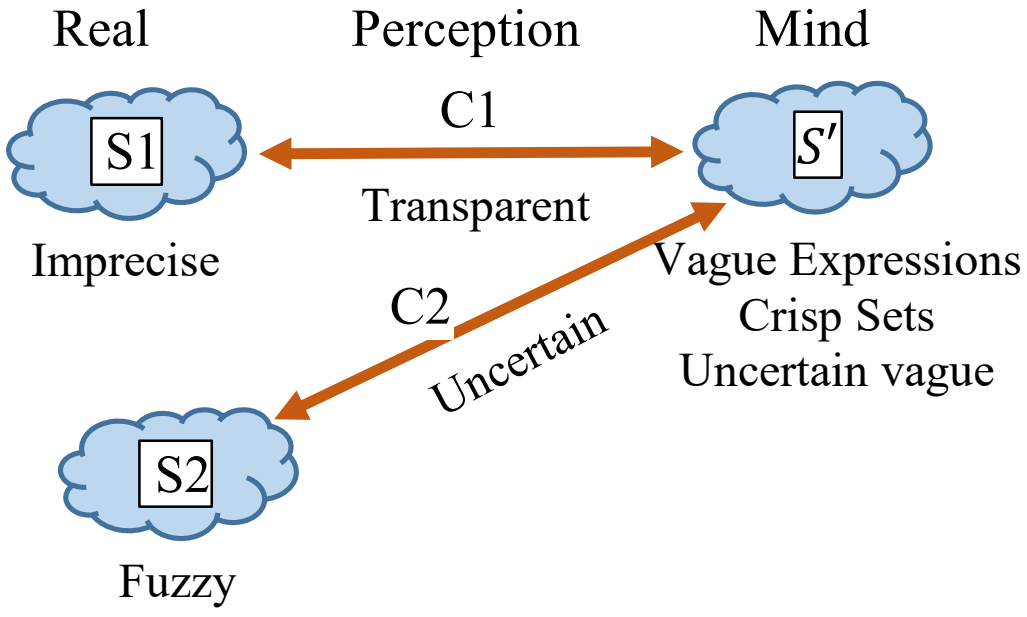 |
| Information Fusion | Imprecise/Uncertain/Fuzzy/logical data sets in real-world<br>Channel has no error<br>Combination rules to Fuse data sets | 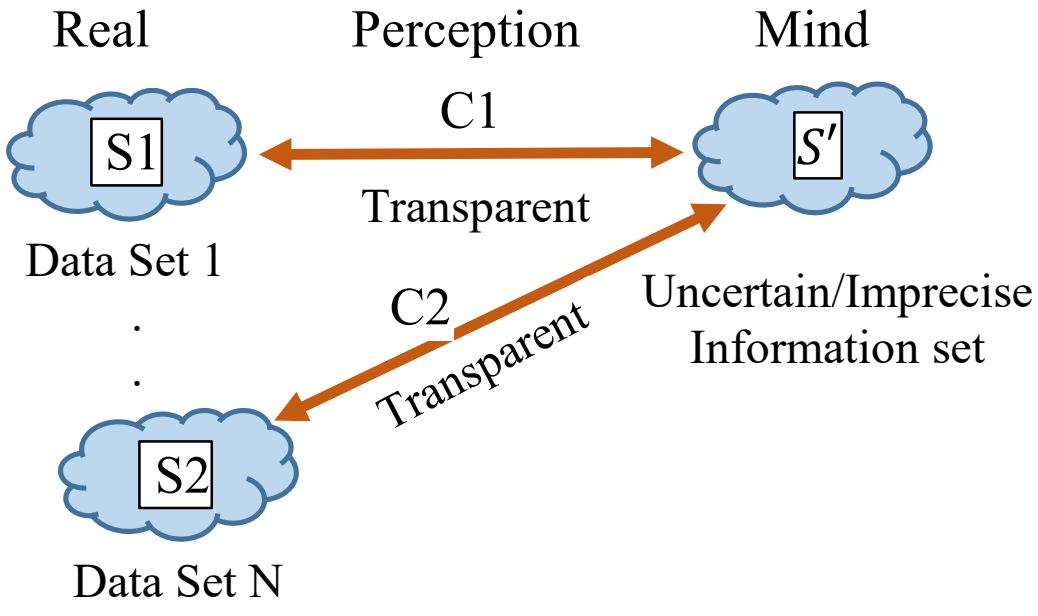 |
| Propositional Logic | Incomplete data in real-world<br>Incompleteness in model<br>Incomplete data in mind<br>Channel model with determined error<br>Channel model with determined data loss | 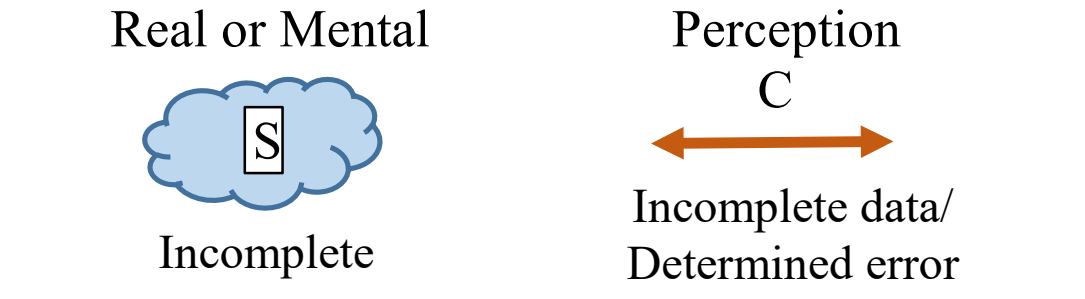 |
| First-Order Logic | Incomplete data in real-world<br>Incompleteness in model<br>Incomplete data in mind<br>Channel model with determined error<br>Channel model with a determined data loss | 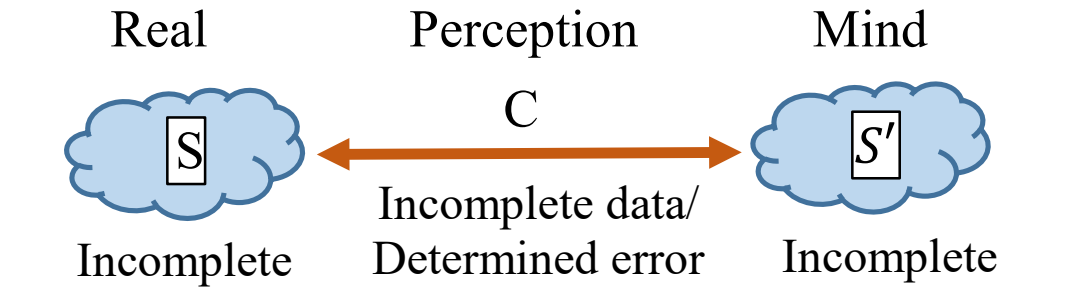 |
| Default Logic | Incomplete/Uncertain situation in real-world<br>Incomplete situation in mind<br>Channel model with determined error<br>Channel model with a determined data loss<br>Default Rules | 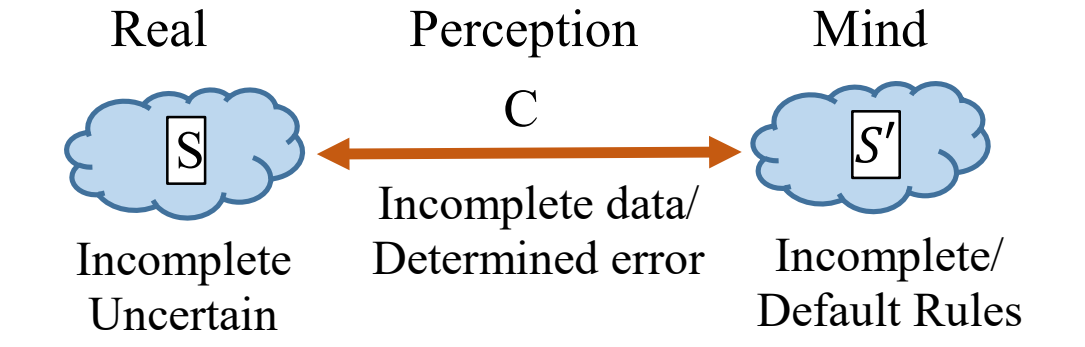 |
| Probabilistic Semantic<br>Probabilistic Logic | Uncertain situations in real-world<br>Uncertain situation in mind<br>Channel has no error<br>Default Rules | 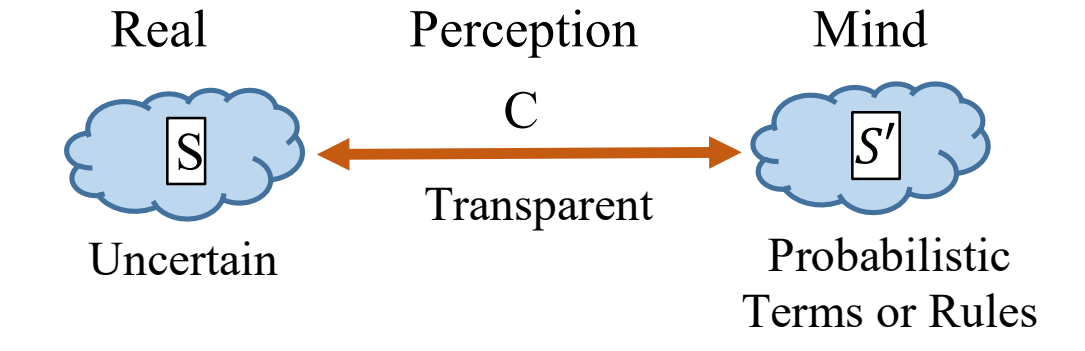 |
| Modal Logic | Incomplete data in real-world<br>Knowledge/Belief state in mind<br>Channel model with determined error<br>Channel model with determined data loss | 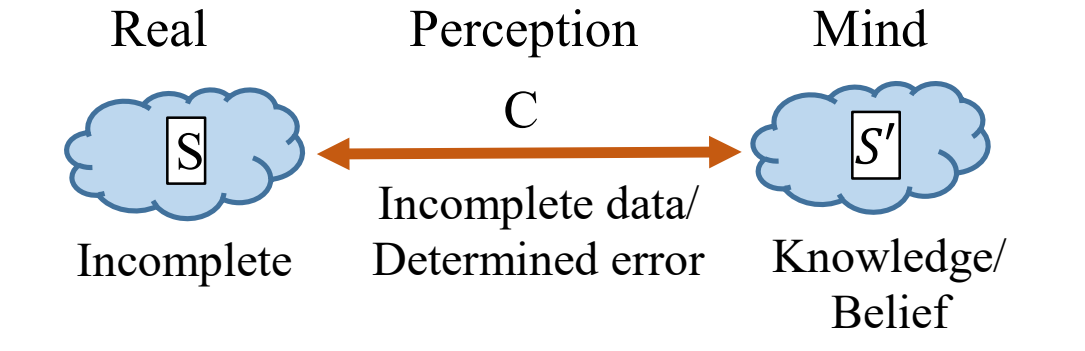 |
| Probabilistic<br>Modal Logic | Uncertain situations in real-world<br>Uncertain situation in mind<br>Channel has no error<br>Default Rules<br>Combination rules to consider some sources<br>Probabilistic channels | 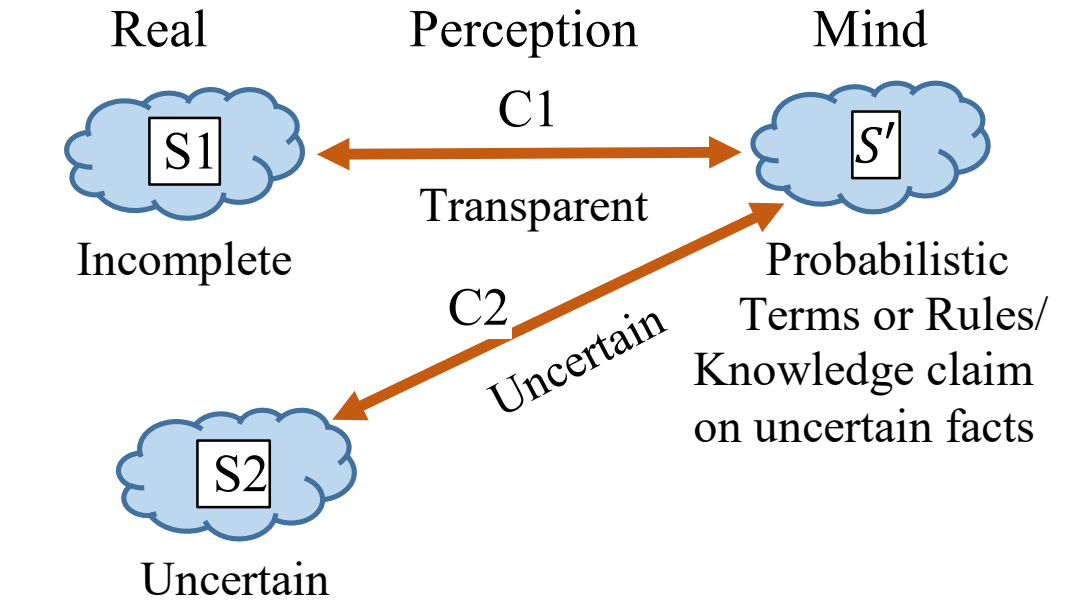 |
| Situation calculus | Context models in real-world/mind<br>Lifting contexts to another context<br>Certain/Uncertain fluent between situations<br>Channel can be for<br>perception/action/translation/.. | 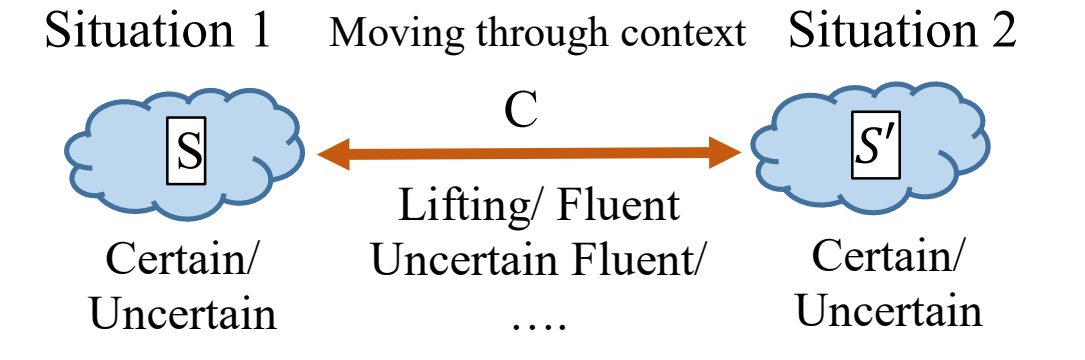 |
| Situation Logic | Incomplete data in real-world<br>Incompleteness in model<br>Incomplete data in mind<br>Channel model with determined error<br>Channel model with determined data loss | 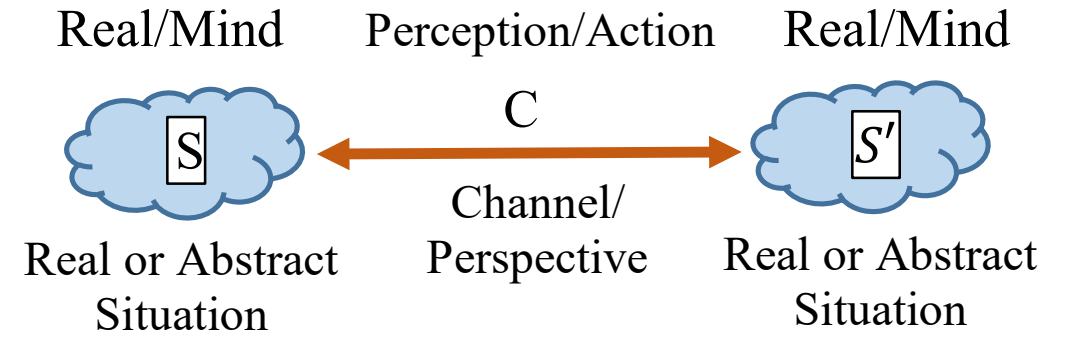 |

*Table 1:* Different Imperfection mangement models and their counterpart in Network Framework

To give more reasoning about my claim, I describe the outstanding feature of Situation theory and Channel theory for imperfect information handling based on the criteria in section 2.2. Also, I start each section by defining the basic concepts of this theory. and introduce more advanced concepts whenever necessary.

## 4.1. Partiality

The most important concept in situation logic is the ***situation***. A *situation* or *real situation* is a partial subset of a possible world, which can be picked out by a cognitive agent. This first definition is the most attractive one, as Barwise showed us that he was not going to rely on a complete set of information or a comprehensive world. It seems that he was going to remind us that we are not in the position of god to be able to access all information. The definition of the *situation* is referring to the basic property of this theory, i.e. ***partiality***.

As I described in previous sections, most models of information have relied on the *possible worlds semantic* that is based on the hypothesis that all possibilities are known to a cognitive agent. This assumption is not true in situation theory; Situation semantics rely on partial subsets of the world or situations. This duality is in the different semantics that has constituted the basis of different models of information. To me, those models that don't have a clear picture of the problem, have mainly tried to construct their theory on the possible world semantics. I cannot imagine modeling the information in the world and mind uniquely when we have supposed a universal set for their entities.

There are two main approaches to model information in systems, Kripke Possible World Semantic and Barwise Situation Semantic. A possible world is a function from all space-time points to a set of possible states for each point. A possible world represents a possible state of affairs in the world. A situation or real situation, as defined earlier, is a partial subset of a possible world, which can be selected by a cognitive agent. Kripke semantic, which is based on the concept of possible worlds, has a global approach to looking at a system. Barwise believes that Kripke's semantic takes "God's eye view" (Barwise et al. 1993). From the system theory point of view, according to Kripke semantic, it is assumed that there is one single "state space" for the entire system. However, the situation theory has a local approach to looking at a system. It assumes that the global view is not possible because of the psychological limitations of any cognitive agent. Contrary to the global view, Situation theory postulates system theory based on a partial understanding of the system and gradually developing and generalizing that understanding.

Accordingly, I think the underlying semantic for a general framework in imperfect information management could be the Situation Theory in which a partial and evolutional concept of knowledge and information has been developed. Imperfect information is the result of partiality and every system that wants to model imperfection, must be consistent with the partiality assumption in the system.

## 4.2. Ontological and Epistemological Uncertainty

A clear distinction between ontological uncertainty and epistemological one is obtained if we could model both real-world information and mental representations. To see the method of situation theory to do this let's review some definitions.

The second important concept particular to this theory is ***individuation***. *Individuation* is a scheme for discriminating, classifying, and objectifying atomic entities including, *objects*, *properties* they might have, *relations* in which they might stand to other objects, *locations* they are placed, *times* they are in, *truth-values* the relations have, and so on. Every cognitive agent, like humans, makes the basic classification conversion of perceived information to individuate or objectify the above atomic entities. So to speak, *individuation* somehow helps us to reach a statistical or fuzzy definition of the entities and relations of the world. So it has a numerical and approximate nature we like in this article.

Let me explain why this notion is unique and important in the theories of information and cognition. This notion is the main tool for the evolution of the theory. Most previous theories are based on static objects and relations that must be identified by the expert that is going to use them. But situation theory accepts just *situations* as the basic things in the theory and insists on exploring or finding other objects based on induction on *situations*. This concept is augmented and formulated better in the channel theory. The concept of individuation turns to the concept of classification in channel theory. In the framework of channel theory, every theory can be modeled by a classification. I will return to this notion in section 4.1. for defining the *Channel*.

It is now possible to introduce the third important concept of situation theory, i.e. ***infon***.

**Definition**: An *infon* is the smallest part of information. It is a simple fact about a situation that is represented as:

$$\ll \boldsymbol{R(a_1, a_2, \dots, a_n), l, t, i} \gg \qquad (4.1)$$

The first part is called a state of affairs, which shows objects $\boldsymbol{a_1, a_2, \dots, a_n}$ stand in the relation $\boldsymbol{R}$. $\boldsymbol{l}$ denotes a spatial location, and $\boldsymbol{t}$ denotes a temporal location. $\boldsymbol{i}$ is called the polarity of an infon. The above expression says that the relation *R* does (does not) exist among elements $\boldsymbol{a_1, a_2, \dots, a_n}$ in location $\boldsymbol{l}$ and time $\boldsymbol{t}$ depending on the value of $\boldsymbol{i}$ that could be **1**(or **0**).

**Definition**: We write $\boldsymbol{s \vDash i}$ to mean that the situation $\boldsymbol{s}$ supports infon $\boldsymbol{i}$.

For example, suppose $\boldsymbol{i}$ is the information that "I am driving". This is supported by the situation $\boldsymbol{s}$ that I am currently driving in it:

$$\boldsymbol{i} = \ll \boldsymbol{Drive(Bob, car), l, t, 1} \gg$$
$$\boldsymbol{s \vDash i} \qquad (4.2)$$
$$\boldsymbol{s} \vDash \ll \boldsymbol{Drive(Bob, car), l, t, 1} \gg$$

The notion of the *situation* in situation theory has two kinds. One is the *real situation* and the other is the *abstract situation*. *Real situations* refer to a subset of the real world mainly restricted to a special location-time region. Situation theory models real-world information based on the individuation and objectification process in real situations. Real situations are the corpus of data to be carved up by cognitive agents to generate *infons* or information.
Abstract situations are mental and abstract forms of real situations in the minds of cognitive agents. The information of a cognitive agent about a real situation could be represented by a subset of *infons*. These *infons* are built by those objects and relations that have been individuated from the real situation.

**Definition:** An *abstract situation* $\boldsymbol{s'}$ that consists of all *Infons* supported by real situation $\boldsymbol{s}$ is the mathematical counterpart of that real situation, which is constructed by a cognitive agent:

$$\boldsymbol{s'} = \{\boldsymbol{i} |\ \boldsymbol{s} \vDash \boldsymbol{i}\} \qquad (4.3)$$

Beware that this set could be non-well-founded if $\boldsymbol{s}$ is a circular situation, i.e. referring to itself.
To apply situation theory to model cognition in a system, we need to classify situations to be able to pick out features that those situations may or may not have in common. That is to say, we need to apply an *individuation* scheme to extract higher-order uniformities, called *types*, across situations. In other words, the basic information, which is supported by a situation, determines how the situation is classified, i.e. how it is *typed*.

**Definition:** If situation $\boldsymbol{s}$ supports the information $\boldsymbol{i}$, written $\boldsymbol{s} \vDash \boldsymbol{i}$, then it is of *type* $\mathbf{x} = \{x | x \vDash \boldsymbol{i}\}$.

The *types* are mental entities that are abstracted by cognitive agents from situations. Both abstract situations and types are mental objects in the mind of a cognitive agent. So situation theory does model both, the world and mind altogether; or has both ontological and epistemological perspectives to information.

## 4.3. Belief and Knowledge

As noted in previous sections, a good framework for imperfect information management must be able to represent two mental states belief and knowledge to discern reliable information from unreliable one in the mind of the cognitive agent. He is going to represent reliable true information as knowledge. But the information that is not reliable and cannot trustfully be considered truth as just some belief.
The other concept in situation theory, which is closely related to the concept of knowledge, is the ***proposition*** that is defined based on the pairs of *infon-situation* as below:

$$\boldsymbol{s} \vDash \boldsymbol{i} \qquad (4.4)$$

This proposition means Infon $\boldsymbol{i}$ is an item of information that is true of situation $\boldsymbol{s}$. The official terminology is that "$\boldsymbol{s}$ supports $\boldsymbol{i}$". This definition reveals the contextual nature of knowledge. In other words, knowledge is “information in context”.
When a cognitive agent acquires a proposition ($\boldsymbol{s} \vDash \boldsymbol{i}$), he knows the information $\boldsymbol{i}$ in situation $\boldsymbol{s}$. This point leads us to the definition of *knowledge* in situation theory, which is based on the Dretske theory of epistemology:

***Knowledge is a belief that is based on information.***

Suppose you have a belief $\boldsymbol{b}$ about a situation $\boldsymbol{s}$. If you succeed in relating that belief to the respective situation, you actually will know ($\boldsymbol{s} \vDash \boldsymbol{b}$). It must be noted that the acquisition of this relationship needs an active involvement in the respective situation. Knowledge is acquired through the experience of or active involvement in different situations. From this point of view, imperfect information handling itself is also a process of capturing knowledge from our partial belief by active engagement in related situations.

## 4.4. Uncertainty and Imprecision

As I noted in previous sections, the nature of incomplete information comes back to our limited perceptions of the world. Statistical evaluation of these perceptions emerges in probability measures in numerical approaches and also emerges in default rules in symbolic approaches. The usefulness of situation theory is in the introduction and formulation of *individuation* as an underlying mechanism to achieve such higher abstractions to grasp and model *uncertain* information.
Another feature of individuation discussed in the literature is the somehow fuzzy or analog to digital nature of it. This is necessary for a cognitive agent to reduce the amount of information in its mind. Also, it is inevitable considering the limited precision of cognitive agents’ sensors. This way, situation theory has accepted and formulated the *imprecision* or *fuzziness* in its theory as a real and necessary part of world-mind interaction. To clarify the usefulness of situation theory in modeling imprecision and uncertainty, I will refer to an example of it in the realm of Information Retrieval systems and then continue with channel theory for a more complete model of both *uncertainty* and *imprecision*.
First, I need to introduce another basic notion of situation theory as below:

**Definition:** A **C*onstraint*** *is a relation between situation types as below:*

$$\boldsymbol{S} \Rightarrow \boldsymbol{S'} \quad \text{(read } S \text{ involves } S'\text{) iff every situation } \mathbf{s} \text{ that is of type } \mathbf{S} \text{ is of type } \boldsymbol{S'} \qquad (4.5)$$

C*onstraints* provide the situation theoretic mechanism that captures the way that agents make inferences and act rationally. Constraints are linkages between situation types. They may be natural laws, conventions, logical rules, linguistic rules, empirical, law-like correspondences, etc.
Situation theory has been used in information retrieval (IR), where the concept of *conditional constraint* has been used to represent uncertainty. (Referred in review by Abdulahad et al, 2019). In that research, each document is represented by a situation "***d***" and necessary queries to search in documents are represented by some infons "***q***".
The sentence $\boldsymbol{d} \vDash \boldsymbol{q}$ or "***d*** support ***q***" means that document ***d*** is relevant to a query ***q***. This sentence is used as a retrieval decision on documents.
The uncertainty related to this decision is estimated based on conditional constraints:

$$U(d \vDash q) = \begin{cases} 1 & if \quad d \vDash q \\ max_{\{D'|(D \Longrightarrow D'|B),,\exists d':D',d' \vDash q\}} \ \delta(D, D') & otherwise \end{cases} \qquad (4.6)$$

Where ***D*** is the type of ***d*** and $\boldsymbol{D'}$ is another type related to ***D***, under some condition ***B*** and $\delta(D, D')$ is defined as below:

$$\delta(D, D') = \begin{cases} 1 & if \ D \Longrightarrow D' \\ 0 < a < 1 & if \ D \Longrightarrow D'|B \\ 0 & otherwise \end{cases} \qquad (4.7)$$

A query ***q*** on document ***d*** is relevant if it has supported by the situation ***d*** itself or if it has supported by another document $\boldsymbol{d'}$ of type $\boldsymbol{D'}$ which is a constraint to ***D***. The degree of relevance of ***d*** to ***q*** can be determined by the relevance between document ***d*** and another document $\boldsymbol{d'}$ which support the query ***q***.
This type of reasoning based on the uncertainty of relevance in IR systems is somehow representing the power of situation theory in imperfect information management. The degree of uncertainty in situation theory can be represented and manipulated not only by examining related situations and infons, but also by examining and manipulating the types of related situations and infons.
For a better understanding of the power of situation theory in modeling information, I first continue by extending the framework that I gave in the introduction of the article. As I mentioned, the imperfection problem can be modeled by considering the model of the world as a set of propositions on which there is a probability distribution or a logical structure. Based on the previous text on different imperfection models, I add another such model for the mind of cognitive agents. This way, the uncertainty in the world and the cognitive agent mind are separated from each other.
The new definition that I want to add to this framework is the notion of the ***channel*** as proposed in situation theory. There is a third thing that relates the mind of the cognitive agent to the world and it is the ***channel*** on which it can grasp information. There are different kinds of channels for different kinds of information, like language for receiving lingual information, vision for receiving images, and different communication media to transmit different kinds of files containing different kinds of information.
To show the overall model, I start by defining ***Classification*** as below:

**Definition**: A ***classification*** $A = \{ \text{tok}(A), \text{typ}(A), \ \vDash_A \}$ consists of a set ***tok(A)*** of objects to be classified, called ***tokens*** of **A**, a set ***typ(A)*** to classify the objects called the ***types*** of **A**, and a binary relation $\vDash_A$ between *tok(A)* and *typ(A)* to determine which *tokens* are classified as which *types*. "$a \vDash_A \alpha$" represents "a" being of type "$\alpha$".

A familiar example of classification is that of a probability distribution. Consider two *types* EVEN and ODD for different events of a roll of a dice. Here, the set of all outcomes of the dice roll is the *tokens* of this classification. Then $r_i \ \vDash EVEN$ if roll $r_i$ has been equal to 2,4 or 6.
This way we can model probabilistic-based issues by classification concept. In general, there is no need to provide the necessary details about how to derive the probability distribution function, based on the relation between *token* and *type* of classification.
For another example, we might classify the height of different persons using three *types*: TALL, SHORT, and MEDIUM. For example, if $p_i$ is a person, then $p_i \ \vDash TALL$ if $p_i$ is a tall person based on some definition (say, based on some fuzzy set distribution).
As a logical example, let me refer to first-order logic theory ɸ. Here the tokens are first-order mathematical structures $\omega_i$ and types are the set of well-formed formulas in ɸ. Then $\omega_i \vDash_\Phi \varphi$ if formula $\varphi$ is true in model $\omega_i$.
Also, imprecise information as modeled by the Rough set theory can be demonstrated by an evolutional classification as in work of Skowron, et al (2018).
The above examples show the power of the notion ***classification*** to model different kinds of information like probabilistic, fuzzy and logical in a general framework raised from situation theory.
The next step is to model the relation between the agent's mind and the real world. Let me first check this for probabilistic information. To me, the first best model of this relation is the notion of "***Channel***" as defined in channel theory in communication engineering and first proposed by Shannon (1964). Let me give an example in probabilistic modeling. If I see the model of the world as a probability distribution and the model of the mind also be another probability distribution, then the relation between these two models could be naturally considered as a channel based on the Shannon theory. As in the channel theory of Shannon, the receiver probability distribution function of received messages is a function of both transmitter probability distribution and channel probability distribution as below. I will give some definitions below to reveal this relation:

**Definition:** We define a *discrete channel* to be a system consisting of an input alphabet X and output alphabet Y and a probability transition matrix **p(y|x)** that expresses the probability of observing the output symbol *y* given that we send the symbol *x*.

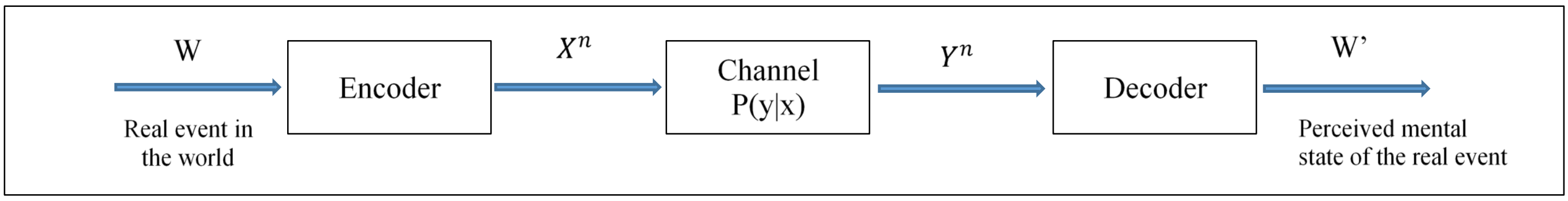

***Figure 3***

H(X) or entropy of the events probability distribution in the real world is representing the intrinsic uncertainty in the information in the real world that the cognitive agent is going to perceive. For a determined and occurred event, the entropy measure is 0 and there is no uncertainty about the real-world state of affairs. On the other hand, if there are other possibilities for a real-world event, then the entropy measure is not zero and there is uncertainty about the actual state of affairs in the real world. As in the literature of information theory, H(X) is a measure representing uncertainty in the events governed by a probability distribution.

- Dice is thrown and side "1" is up → The entropy is zero → No uncertainty
- Dice is not thrown → The entropy is not zero → There is uncertainty

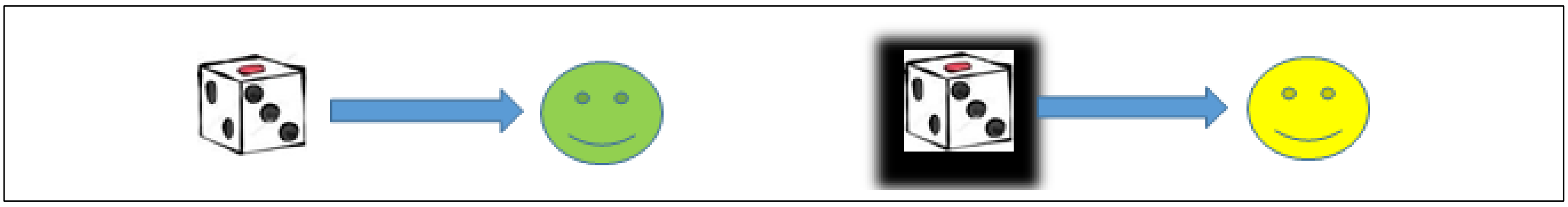

***Figure 4***

In this article, I am not going to give a unique measure for uncertainty and my example about uncertainty based on entropy is just a first step to take and then show the next advantages of taking channel theory interpretation for the process of cognition. A detailed investigation of Shannon's theory from the cognitive point of view is the astonishing work by Dretske (1981). Dretske founded a new theory for Epistemology based on the notion of information and he proposed some new interpretations of Shannon's theory in his investigation. Frankly speaking, Shannon represented information and entropy by an average measure to reach his own goals in engineering: famous source coding and channel coding theorems. But, Dretske tried to define the notion of information in such a way that can be used to represent a single event for his own goal: his famous definition of knowledge based on belief and information.
Now, let me take the second step in my reasoning about uncertainty modeling. Cognitive agent perception of the real world could be different from the real-world state because of the second part of the channel model of the cognition, i.e. channel itself. Again, I give another simple example to show the share of this part in overall uncertainty in the cognitive agent's mind.
Imagine that a real event in the world has occurred (real world with 0 uncertainty) and the cognitive agent is trying to find it by its sensors. Unfortunately, his sensor (Or channel) has some errors and the received information could have some errors based on a probability distribution function. Then, again there is not zero entropy in the received information. The probability distribution of different events at the receiver (Or cognitive agent's input) is exactly the probability distribution of errors in the channel.

- Dice has thrown and side "1" is up → The entropy is zero → No uncertainty
- An opaque obstacle in front of the dice → Entropy is not zero → There is uncertainty

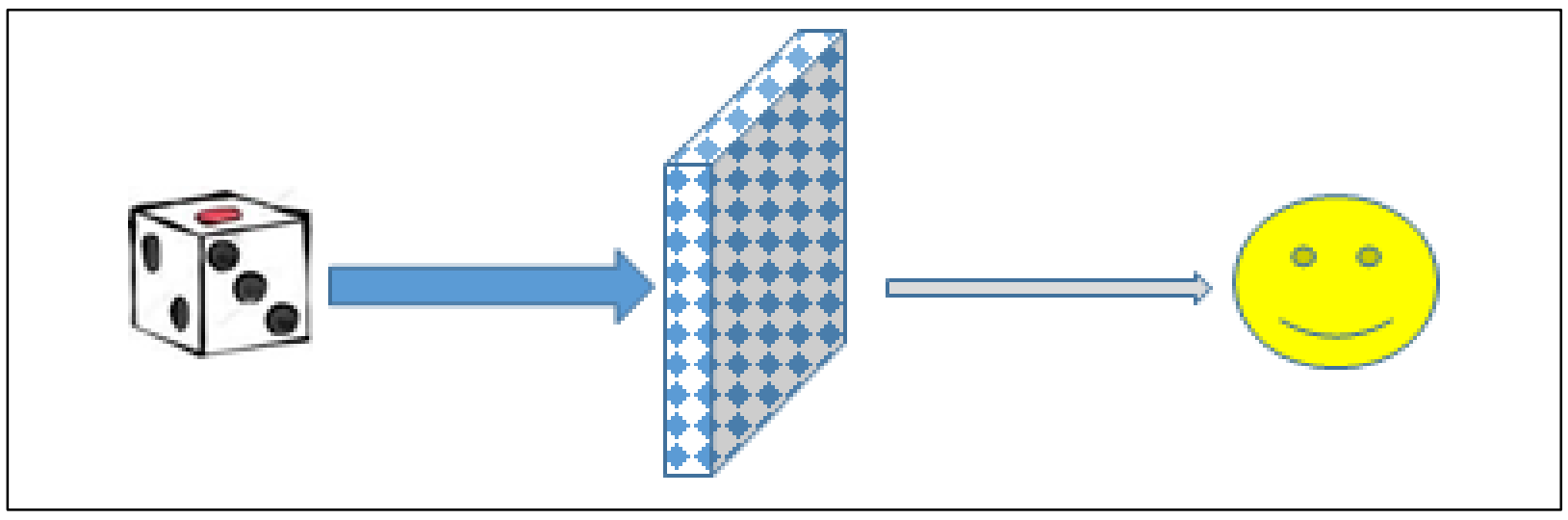

***Figure 5***

Up to now, I considered the effects of real-world uncertainties and channel-made uncertainties. Let me now refer to the last source of uncertainty based on the channel model of cognition. Imagine that there is no uncertainty in real-world events and also in the channel, but the cognitive agent itself perceives the received state of affairs and put it in a less accurate model than it was, or understands it with less powerful language. For example, imagine a cognitive agent that knows just the number "1" and considers other numbers as "non 1". To him all numbers of dice are just "1" or "non 1". This might happen because of his strategy or maybe because of his cognitive deficiencies. Although this is very strange, I remember reading about a very strange tribe in Africa in which their people discriminate all colors by two notions in their language. More research then showed that the simplicity in their language affects their perceptions in such a way that they can see just two colors in everything and no more.

- Dice has thrown and side "2" is up → The entropy is zero → No uncertainty in source
- Channel is error free → No uncertainty in channel
- Cognitive agent ability is low → Increase in entropy → There is imprecision

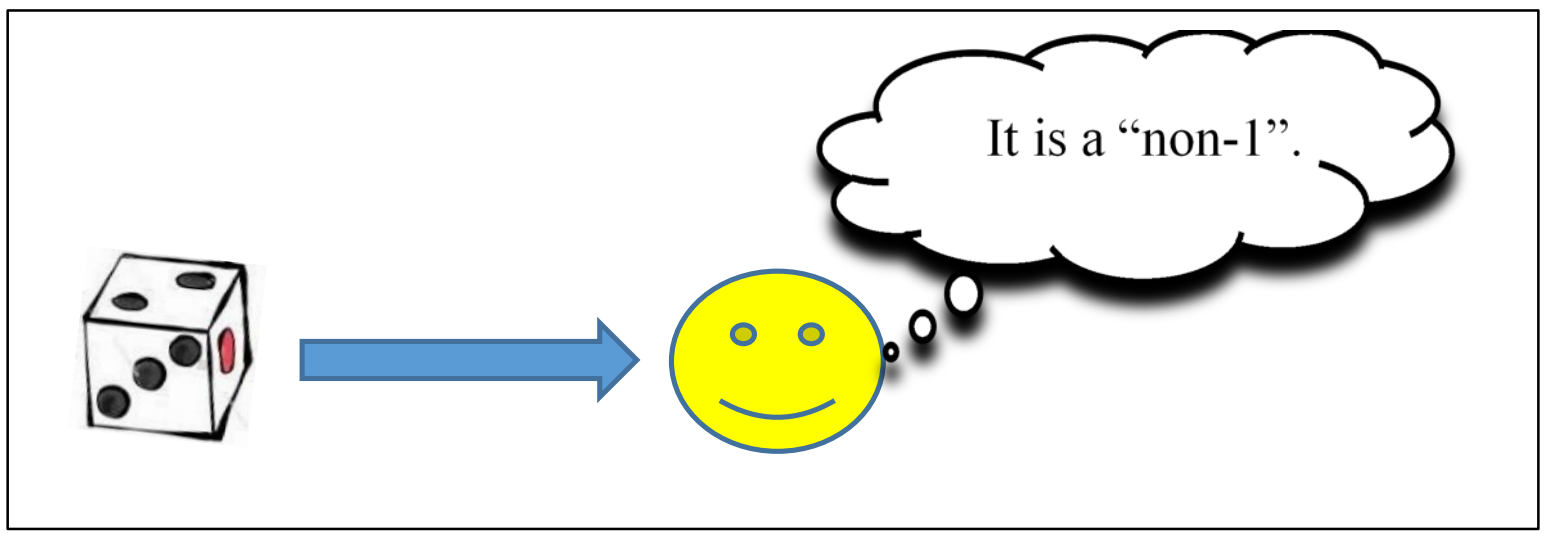

*Figure 6*

I had better note that this kind of imperfection can be modeled by channel in some cases in which perception layers of cognition are responsible for it. This is an example of modeling ***imprecision*** in cognitive agent perception. In another scenario, the cognitive agent is going to inform another cognitive agent in a simple language about the result of his observation like the below figure. As we know, the lingual representation of ***imprecision*** is ***vagueness***.

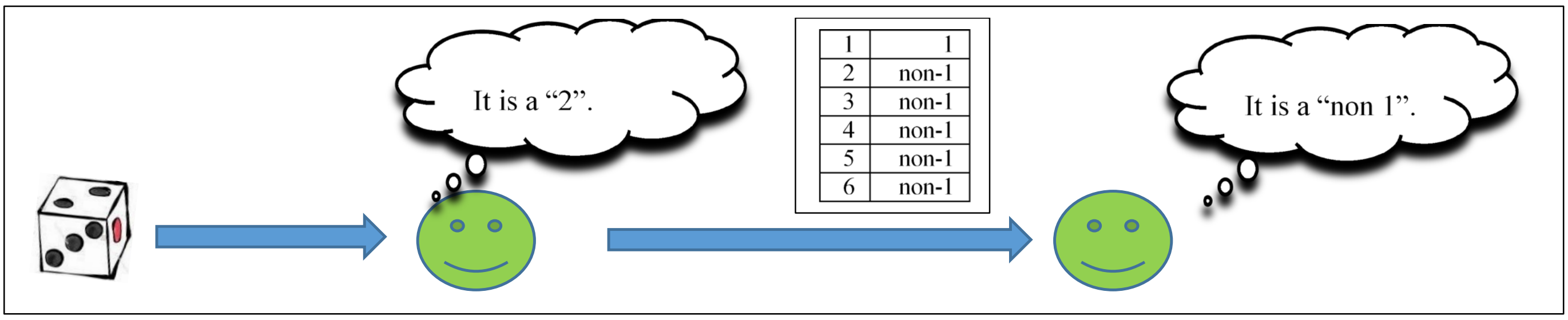

*Figure 7*

All the above examples based on Shannon's theory of channels show us the powerful descriptive abilities of channel theory. Shannon's theory of channels is based on probability theory. As noted in previous sections, imperfect information management must include logical models too. To reach this goal, I must rely on the generalization of channel theory as performed in Situation theory as proof for my claim in taking this theory as a general framework for imperfect information management.
The basic concept to depict this image is that came first from situation theory and then continued in the "information flow theory" of Barwise and Seligman (1997) and also completed then by Seligman (2009).

**Definition**: Let $A = \{\mathrm{tok}(A), \mathrm{typ}(A), \vDash_A\}$ and $B = \{\mathrm{tok}(B), \mathrm{typ}(B), \vDash_B\}$ are classifications. An infomorphism is a pair $f = \{f^{\wedge}, f^{\vee}\}$ functions as below:

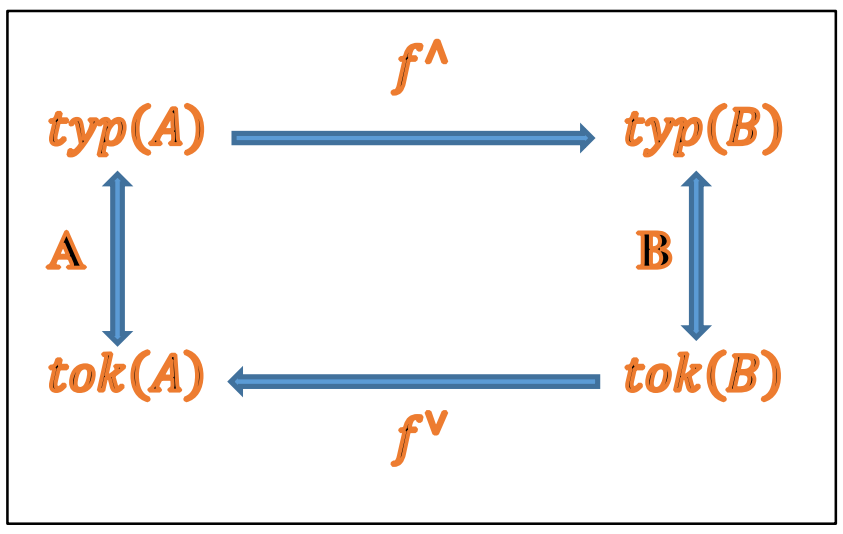

*Figure 8*

And satisfying $f^{\vee}(b) \vDash_A \alpha$ *iff* $b \vDash_B f^{\wedge}(\alpha)$ for all *tokens* b of B and all *types* $\alpha$ of A. This infomorphism can be represented by $f\colon A \rightleftarrows B$.

Definition of Infomorphism as a relation between classifications is used in situation theory and also distributed system theory to grasp the notion of information flow. The existence of informorphism between two classifications A and B shows that they can carry information about each other both in *types* and *tokens*.

**Definition:** A (binary) channel from classification **R** to classification **M** consists of a classification **O** and infomorphisms $f\colon C \rightleftarrows R$ and $g\colon C \rightleftarrows M$. Classification **C** is called the core of the channel:

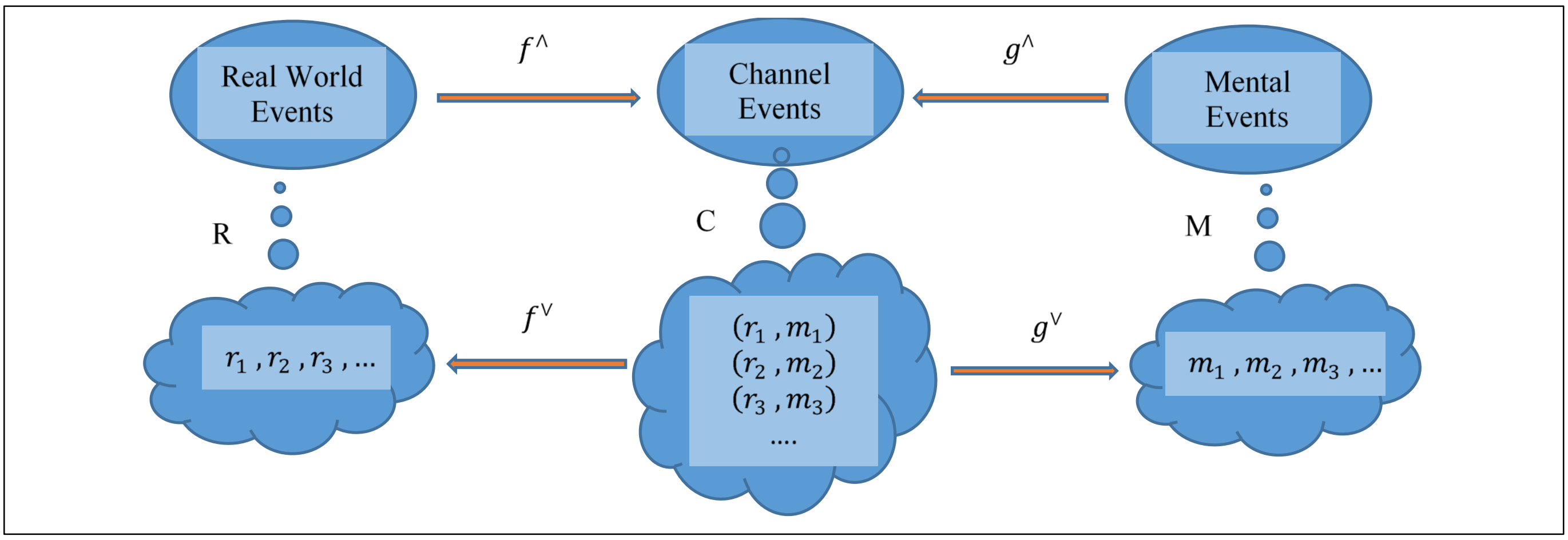

***Figure 9***

A good example of the channel as defined above is the notion of the Shannon channel as described earlier to represent different types of uncertainty when a cognitive agent is going to perceive a situation. The generalization of channel definition as above gives us a more descriptive ability to cover all kinds of information that we need for the aim of this article as described in follow.

## 4.5. Unique Information Theory

As mentioned earlier, to reach a unique and general framework of imperfect information management needs us to have a unique theory of information. I am going to use it in our problem to show that there could be a unique framework in information modeling to support the main question of this article. The notion of the channel as defined above can cover both the logical concept of information and the probabilistic one. I considered the real world as the sender **S** and the cognitive agent as the receiver **R** and the process of understating the world as a channel **C** between them.

From a probability theory point of view, information is in those events that could decrease the entropy of the probability distribution assigned to the events of the probability space of **S** or **R**. Information could decrease the number of possible worlds or possible outcomes in the universe of discourse in probability space of **S** or in probability space of R. The cognitive agent has assigned that distribution and try to decrease the uncertainty about the actual occurred event. Also, as much as we decrease the error-like conditions in channel **C**, the uncertainty that cognitive agents receive will be decreased. How much we succeed in eliminating unrelated possible tokens from the set of tokens in all three parts of the channel, the fact will transfer more and more clearly. This approach to information is based on probability and has been developed well by Shannon and completed by Dretske from an epistemological point of view.

Another notion of information we need is based on logic. One of the main strands of modeling in imperfect information management is logical ones that I had a survey in section 3. Information from a logical point of view is in those propositions that could help cognitive agents reason logically and reach the correct conclusion about the actual state of affairs in the world. More information means more propositions about the real world and it results in new valid conclusions. In this framework, we can study different kinds of uncertainty in different parts of the channel as I did for the probabilistic approach.

As revealed from these explanations, both probabilistic information and logical information can be modeled by channel theory. Handling imperfect information in an integrated manner needs a unique theory of information. Fortunately, this has been propounded by Barwise in his theory of situations and formulated by Seligman in 2009 in his article: Channels: from Logic to Probability for this purpose.

Seligman showed that these two notions of information, as he named "via Logic" and 'via probability" are equivalent based on the notion of the channel in situation semantic. This work is mainly on adapting the Barwise-Seligman framework (1997) to provide a similar account of information, both probability-based and logic-based. He named probabilistic type channels as "Shannon channel" and logical type channels as "Tarski channel". The theorem that Seligman proved gives us the confidence that situation theory plus channel theory is a suitable general framework in imperfect information management for both probabilistic and logical modeling as reviewed in sections 2 and 3. His fundamental theory has many preliminary theorems and definitions that cannot be summarized here. The important point for my research is to comfortably use channel theory both for numerical and symbolic information based on that unifying theorem proved by Seligman.
It is better to note another close classification of information here. In the work by Benthem et al (2008), three major accounts of information have been named “information as range”, “Information as code”, and “information as correlation”. These notions of information are another expression of those I gave above. “Information as range” is a semantical notion of information based on possible worlds. More information is in those events that decrease the range of options. This is another expression of the definition of information as the “reduction of possibilities” in Barwise et al (1997) and has developed well by numerical methods like probability theory. “Information as code” refers to those premises that lead the inferences process toward specific conclusions. It is the same notion of information as I said in symbolic approaches. The “information as correlation” refers to that content that is in a situation about another one. Benthem has identified three types of information but the channel theorem of Seligman has shown us that the third type as in Benthem’s classification is the unifying type of the first two.

## 4.6. Vagueness and Sorites paradox

As I noted in sections 3.1.4 and 3.1.5, Fuzzy models of information have great roles in imperfect information handling. The models of fuzzy set theory and possibility theory have a remarkable influence on AI and logic. But, there are a few problems that must be resolved, two of them are as follows:

- The first one is the Sorites paradox as I mention in 3.1.4 which prevents us from being comfortable using fuzzy and vague concepts in our reasoning.
- Another major problem is the possibility of plausible reasoning and communication of cognitive agents in a shared situation.

Situation theory and channel theory have reasonable answers to these questions as presented in Barwise et al (1997). I give a simple presentation of this approach with my own interpretation here.
The first step is to define vague predicates as classifications. Barwise used the vague concept “height” as an example as below:

- B is the set of some physical objects.
- SHORT(X), MEDIUM(X), TALL(X), TALLER(X,Y), and SAMEHT(X,Y) are formulas for describing the concept types.
- A is the set of all variable assignments taking values in B for the above formula.
- Tok(A) is the set of all tokens of this assignment.
- Typ(A) is the set of all types of this assignment.
- The classification Σ = {typ(A), tok(A), $\vDash_A$ }, $\vDash_A$ is the assignment of tokens from B to some properties in types.

As an example of the assignment:

- $a \vDash_A TALL(X) \wedge TALLER(Y, X)$ means that object a(X) assigned to X is Tall and a(Y) is taller than a(X).

This formulation of the vague concept “height” is not unique and each cognitive agent can have its dedicated classification. The central mathematical object that can help cognitive agents communicate with such different classifications is the definition of suitable state space for the “height” concept to be used as a channel.
**Definition**: Event State-space or Evt(S) of a vague concept is defined as below:

- **Ht** is the state space with tokens in **B** and types or states as the real number representing the height of the tokens.
- **S** is the state-space with tuples of tokens in **B** and tuples of real numbers as heights of those tokens.
- Evt(S) is the state space with tuples of tokens in **B** and a set of formulas as types that are true for those tokens.

Evt(S) is a standard classification that assigns to each set of tokens a set of formulas that are true for those tokens. This standard classification is used as a channel between different agents that have different classifications. The correspondence between an Evt(S) and a classification created by a cognitive agent is also based on a new notion in channel theory for vague concepts:
**Definition:** A regimentation consists of an (n+1)-tuple $r =< \epsilon, I_1, I_2, \dots, I_n >$ in which:

- $\epsilon$ is a non-negative real number named tolerance of r.
- $I_1, I_2, \dots, I_n$ are the mutually disjoint interval of real numbers representing each type of vague concept. (For the “height” example, there are three intervals to show the range of height for SHORT, MEDIUM, and TALL people).
- The distance between consecutive intervals is at most equal to $\epsilon$.
- The **Regimentation** concept is used to show the real measures that cognitive agents assign to vague concepts.

The tolerance number $\epsilon$ shows the size of the intervals that act as borders between intervals showing the different fuzzy amounts of height.

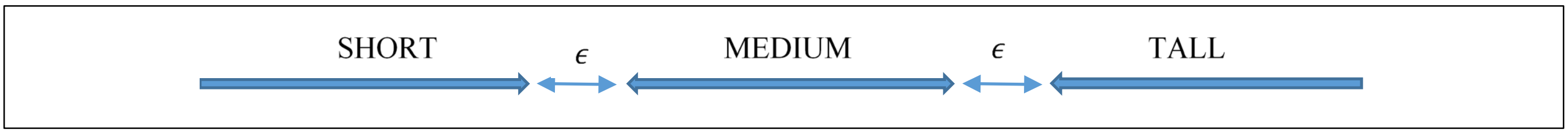

***Figure 10***

The interesting property of Evt(S) is that it is possible to find a regimentation r for each classification A that A and Evt(S) are infomorphism. Simply, each classification of a vague concept like "height" in a cognitive agent can be modeled by Evt(S) by a suitable regimentation r. In the below figure, $f_{r1}$ is an infomorphism from the classification $A_1$ to Evt(S) based on regimentation $r_1$ and $f_{r2}$ is an infomorphism from $A_2$ to Evt(S) based on regimentation $r_2$.

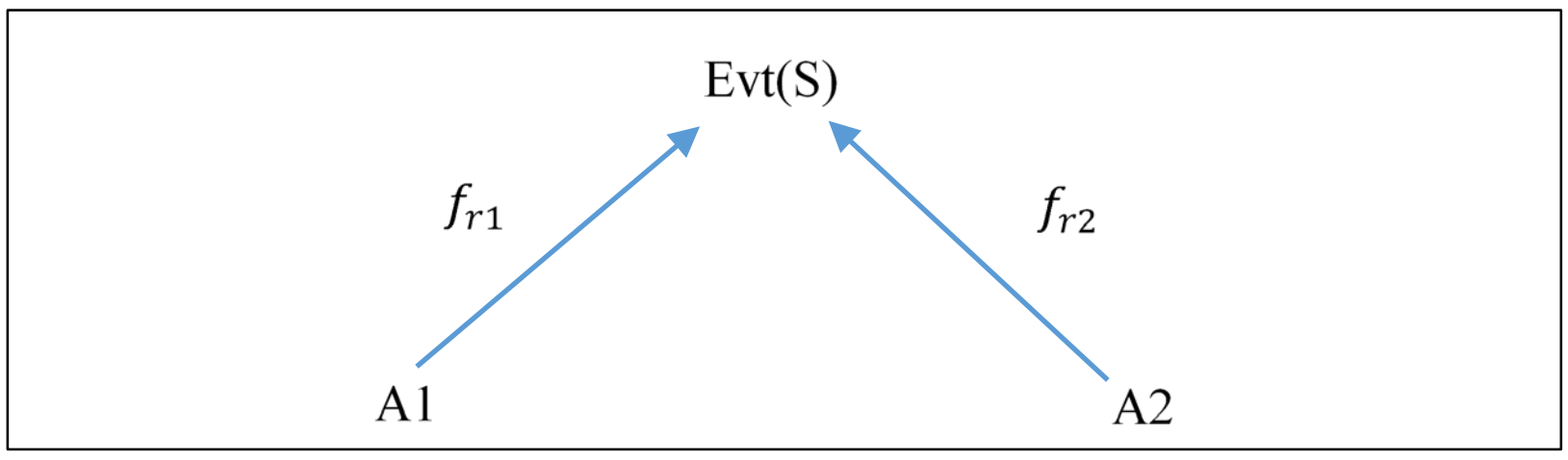

***Figure 11***

In the figure, we can see that there is a suitable infomorphism from each classification of the vague concept in cognitive agent mind to Evt(S) as the standard classification of that vague notion Regimentation. So the Evt(S) can act as a channel between those classifications or those cognitive agents. Selecting suitable regimentation by each cognitive agent for the correct classification of vague concepts makes possible information flow or communication between them. This possibility of information flow between cognitive agents despite their different classifications of vague concepts is the solution to one important problem in this article regarding context dependency.
Another problem that will be solved by this modeling of vagueness is the Sorites paradox. Barwise used an intentional implication of the logic in a classification to solve the Sorites paradox.

**Definition:** A *local logic* $Log(A) = [A, \vdash_{\mathcal{L}}, N_{\mathcal{L}}]$ consists of a classification A, a set $\vdash_{\mathcal{L}}$ of constraint (satisfying certain structural rules) involving the types of A and a subset $N_{\mathcal{L}}$ of tokens of A which satisfy all the constraint of $\vdash_{\mathcal{L}}$.

The local logic of a *classification* is the set of rules that can be seen in the *types* of the classification based on the content inside the *tokens* of **A**. For vague concepts, each cognitive agent has its classification of that concept and therefore has its own *local logic* for that classification. Cognitive agents reason based on their local logic in some cases but usually, they reason based on a minimum theory that is common among different agents because they are also interested in efficient communication with other objects. Based on this reasoning, another concept has been defined below:

**Definition:** *Intensional logic* of A is as below:

$$Log_r(A) = f_r^{-1}\{Log(S)\}$$
$$Log^0(A) = \prod \{Log_r(A) | r \text{ is a regimentation of } A\} \qquad (4.8)$$

The first sentence is representing different local logics derived from **A** that has infomorphism $f_r$ to state-space S. The second sentence or *Intensional logic* of A is the greatest lower bound of these logics having different regimentation. The *Intensional logic* of A includes the basic and common rules for the vague predicate that has modeled my state-space S. based on this logic, the below preposition can be proved for Sorites paradox:

**Theorem**: if **A** is a classification for a vague predicate (Say "height") and is finite (has finite types) then for sufficiently large integer N:

$$SHORT(X_1) \wedge SAMEHT(X_1, X_2) \wedge \ldots \wedge SAMEHT(X_{N-1}, X_N) \nvdash_A^0 \neg TALL(X_N) \qquad (4.9)$$

In which $\Sigma \vdash_A^0 \Delta$ means that $\Delta$ is deduced from $\Sigma$ in local logic $Log^0(A)$. This theorem can be proved by selecting a suitable regimentation 'r' with a suitable tolerance $\epsilon$. Based on the minimum logic that is used in this reasoning, agents have minimum attention to the exact extensional notion of that vague concept (height) and mostly reason based on those types that are common among all agents. So the intensional notion is less accurate and SAMEHT(X,Y) refers to the proximate notion of "having the same height", so consecutive use of that vague relation SAMEHT(X,Y) could lead to a correct conclusion contrary to Sorites paradox.
The key point that Barwise has in this theorem is the use of intensional logic for the reasoning to solve the Sorites paradox. It says that reasoning with vague predicates should be based on a minimal logic for that vague concept to make communication possible and avoid incorrect reasoning like that happened in the Sorites paradox.
The selection of intensional logic could have a psychological interpretation based on the work of Parikh (2019). He has a theory of language based on the situation theory of Barwise and game theory. He also has a solution to Sorites paradox based on psychology. His main reasoning is about determining and interpreting the borderline that is necessary to discriminate a vague concept to its variants. ("borderline" between TALL and MEDIUM and between MEDIUM and SHORT in our example).
He proposed to use a vector of attributes for each vague concept for its modeling. Then using exemplars (or maybe prototypes) of the different values of that concept to reach a probabilistic measure showing the distance between the average of exemplars and a certain token of that vague concept. He argued that it is possible mostly to reason probabilistically which category that token resides, but in borderline between adjacent values, there is a gap that is not decidable. He argued that this borderline gap is not a cutoff line but a distance that is fuzzy intrinsically and is based on human norms to have communication.

## 4.7. Context dependency

In situation theory, it is assumed that cognitive agents have no access to the entire world. They have access to a limited part of the world named situation or context. Each cognitive agent has its individuation of his situation and recognizes its own set of *Infons* of that situation or context., Each agent individuates and understands the world up to his sensory and cognitive abilities. Consequently, the results and conclusions each cognitive agent has in a context could be different from others in the same context. On the other hand, the context itself is a dynamic entity. Every action of cognitive agents or other events changes the context continuously and this, in turn, changes the beliefs of that cognitive agent about that context. Imagine that we consider the agent's interactions and their relative understanding of their common situation. This results in more sophisticated context-dependent beliefs and actions of the agents.
There are some good approaches to context modeling based on situation theory. I introduce some of them here to show the explanatory power of this theory. Situation theory is principally a theory of context (or situation). Situation refers to all contextual facts that could produce information. Inventors of this theory have announced several times that "*information is situated*". Information can be emerged and be identified as an object when we see it in a situation. By this rule, communication is also defined as a relation between situations as we defined earlier. In summary, situation theory is essentially a theory of context that was invented to explain the nature of communication in natural language. Meaning is transferred by understanding the situations of speaker and listener and the relation between these situations.
Although there have been many studies in the literature about the notion of context, most of them accepted it as an essential part of language, and Akman (1997) argued that Situation theory as in Barwise's idea can be applied in context modeling to give us necessary properties of context-dependency like "Dynamic Context" and "non-monotonicity". He believed that the situation theory is useful not only for knowledge representation schemes but also to support contextual reasoning. He used Situation theory for implementing one of the main concepts of McCarthy's theory: "lifting" or "transcendence". He translated the logical reasoning of McCarthy to information-based reasoning by using support relations and constraints. The famous example of {on-above theory} of McCarthy has been used in his article.
Here, for more explanation of context modeling, I demonstrate the concept of situation theory proposed by Devlin (2006) and (2009). Situation theory has concerned with context as a basic concept to provide a relational theory of meaning. The meaning of an expression is defined based on a relation between an utterance or discourse situation (or context), a speaker's connection function, and a described situation (or context). Situation theory is considered a set of mathematically-based tools to analyze, in particular the way context facilitates and influences the rise and flow of information (Devlin, 2006). He proposed situation theory as a model of human reasoning. In his framework, the reasoning is viewed as a temporal cognitive process that acts on entities of the below form as a *basic reasoning element*:

$$S \models_{\tau_1,\tau_2,\ldots} \sigma \qquad (4.10)$$

Where:

- $\sigma$ is a statement or a fact;
- S is a situation that provides support or context of origin for $\sigma$;
- $\tau_1, \tau_2, \ldots$ are indicators of $\sigma$. They are specific items of information in situation S that the agent takes as justification for $\sigma$.

And the process of reasoning (named by Devlin as the *evidential reasoning process*) can be represented:

$$\begin{array}{lcl} s_1 & \models_{\tau_1,\ldots} & \sigma_1 \\ s_2 & \models_{\tau_2,\ldots} & \sigma_2 \\ s_3 & \models_{\tau_3,\ldots} & \sigma_3 \\ & \vdots & \\ \hline\hline s & \models_{\tau_1,\ldots\,\tau_2,\ldots\tau_3,\ldots} & \sigma \end{array}$$

***Figure 12***

Each line in the above process is showing that a fact is supported in a context by some item of information in that context or situation and aggregation of all these elements is an *evidential reasoning element.*
The above element could have different interpretations in different models. For example, by taking situations as a logical structure and taking $\sigma_1$, $\sigma_2$, … as some formula in a first-order language, we have a system for symbolic reasoning. Also, by taking situations as to be equal, then it refers to a reasoning process in a common context. As another example, consider situations as a number between 0 and 1 and interpret the formula in each line as a probability statement $p(\sigma_i) = s_i$, then we have the Bayesian inference system in which final 's' is a numerical function on all $s_i$'s according Bayes' theorem and final "$\sigma$" is the conclusion based on evidence $\sigma_i$'s.
Devlin continues by introducing different rules and operators to entail new *evidential reasoning elements* based on present elements. The difference between Devlin's framework and with the original situation theory is his emphasis on considering indicators. To him, reasoning involves three main components: facts, sources (or supports or context), and indicators. The first two items are in situation theory and the third one that comes from social science is essential for explaining justifications in different contexts.
Another modeling for context in situation theory is in the concept of perspective as defined by Seligman (1990). One of the original motivations behind the development of situation theory was to model visual scenes. A scene is made up of many situations; each one is a visual perception of the scene from a distinctive point of view. The objects, relations, and other individuals in the scene are uniformities across situations of that scene, which can constitute the set of our types, related to those situations by classifying the set of situations.

**Definition:** A *perspective* is a structure $< S, T, \vDash, \Rightarrow, \perp >$ where $< S, T, \vDash >$ is a classification, S is a collection of situations, and T is a collection of types, and $\Rightarrow$ (*involves*) and $\perp$(*precludes*) are binary relations on T, such that for all $s \in S, t \in T$,

- If $s \vDash t$ and $t \Rightarrow t'$ then there is an $s' \in S$ such that $s' \vDash t'$ : *facticity*
- If $t \Rightarrow t'$ and $t' \Rightarrow t''$ then $t \Rightarrow t''$ : *xerox*
- If $s \vDash t$ and $t \perp t'$ then $s \nvDash t'$ : *local preclusion*
- If $t \perp t'$ then $t' \perp t$ : *mutual preclusion*

(4.11)

The *facticity* condition is used to capture the intuition that if someone has some information then that information is a fact. The *xerox* principle is to show that *involve* relation is transitive. The *local preclusion* is to show that a situation cannot have both an information item and its negative information. The *mutual preclusion* is to show that the *preclusion* relation is symmetric.
Generally, a *perspective* is a set of situations about a unique subject in addition to a set of individuated types of those situations. Each situation is acquired by placing the subject in a special *perspective* and each type is a uniformity that refers to the common aspect of the respective situations. This definition of *perspective* is a simplified mathematical definition of the usual concept of perspective and gives us the necessary tools to reason about situations from different perspectives as a special model of context. For a special example of contextual reasoning based on situation theory, you can refer to Barwise et al (1997). They showed how to reason in a situation in a classification about another situation in another classification when they are connected by a channel.

As a short note for less important features of imperfect information management, we know that any rational framework for imperfection management must also take into account non-monotonicity as a specific property in human cognition. Barwise and Seligman (1997) showed the descriptive power of classification theory plus the state-space model of information to model non-monotonic reasoning by the concept of *conditional constraint*. Also, they checked the independence property in the inputs of the state-space model of information to show a solution for the Frame problem.

## 5. Conclusion

I have depicted all the different models that I know up to now in Figure 1. The overall movement in research on imperfect information handling separated from classical logic and probability theories toward those models that have separated the agent's information and knowledge from the real state of affairs. The movement toward new models is based on the fact that humans have many deficiencies in receiving and understanding the facts of the world.
Based on all the previous research that I mentioned in this article and my own reasoning, I propose the below criteria for a suitable framework for imperfect information handling problems. To me, a suitable framework for the management of imperfect information should:

- support **Partiality** in information.
- consider the modeling of the **World facts** and **Mental representation** of them simultaneously.
- consider the mental state of **Knowledge** and **Belief**.
- model **Imprecision** as an essential effect of world-sensors relation.
- model **Uncertainty** in grasped information as a model of sensors or channels performance.
- model **Default rules** in agents' minds and model their numerical counterparts i.e. **event probabilities** in the real world.
- rely on a **unique notion of Information.**
- model **Vagueness** or **Fuzziness** as the lingual counterpart of imprecision.
- model **Context**.
- handle the **Inconsistency**.
- Encompass **Non-Monotonic** reasoning rules.

The situation theory of Barwise and channel theory of Barwise-Seligman is the suitable framework that can encompass all approaches uniquely and clearly. It is originally proposed to model *context* in a *partial* manner. *Partiality* in cognitive agents and *context* are related to each other by the concept of *perspective* or by a rather more powerful concept of a *channel*. Every identified model of information can be represented in situation theory as a *classification*. By considering two *infomorphisms* between a channel and two real and mental *classifications*, one can separate *ontological* and *epistemological* uncertainty and also *sensor uncertainty* in a unique model. They support both *numerical* and *symbolic* models of information by different kinds of classification relations. The support relation in this theory is a suitable generalization of classical logic notions to distinguish the nature of information from knowledge and so could discern *belief* and *knowledge*. Imprecise classification is used in *imprecision* modeling and related channels to that classification toward lingual classification can model associated natural language *vagueness*. *The inconsistency* and *non-monotonic* nature of this theory are derived from partiality and context dependency in it.

The ***network framework*** proposed here is ***open*** and ***scalability***, just like any other networking plan. The concept of scalability can aggregate all research and efforts in imperfection information handling into one unique plan. The open architecture that I proposed guarantees persistent standards and modules that will be developed in this framework. It lets us develop and evolve the system without any severe changes in the previous structure. After accepting this framework, we could get rid of so many difficult modeling of combined approaches. Then a wide and astonishing research area will be started for concentrating on modeling different channels or translations between different information models.

My research is the first step and many kinds of research are required to augment this framework. Although basically, the logical and probabilistic notions of information have unified in this theory, a branch of research is necessary to have similar work for fuzzy information. Information flow through an integrated network of classifications of different information models is another attractive research area that I propose for the practical application of this framework in data fusion.

Throughout my article, I referred to psychological reasoning to support my claim about the cognitive basis of imperfect information issues. This is due to the limitations that are present in cognitive science research on the structure of the brain. Also, I have not referred to the neural network approach as one of the main technics in handling imperfections in information. Although physical approaches in cognitive science have not been developed yet, it is needed to have separate research on the relation between Situation theory and artificial neural networks. I think that there could be important justifications for situation theory in neuroscience. There are two fundamental reasons why they are too close to convincing us for such research. Firstly, both are on a phenomenological basis. As in Barwise et al (1983) and also in Dretske (1981), situation theory was introduced as an ecological theory of semantics. Throughout the origin of this theory, we see many clues to the phenomenological interpretations in dealing with its basic notions of it. Many basic concepts of the situation theory like *classification*, *uniformity, individuation,* and *regularity* are phenomenological concepts. Situation theory is conceived as a developmental theory of information in cognitive agents based on their active engagement with the real world which is one of the main features of artificial neural networks. Secondly, they both refer to the classifying nature of the brain. The notion that has been added to previous theories of information by situation theory is its ability in formalizing the notions like *classification* and *concept formation* that are essential for modeling an evolving cognitive agent. This feature is achieved in engineering by artificial neural network approaches. Artificial neural networks as a simple model of the brain are mainly classifying objects.

So, I think we may obtain strong reasons for using situation theory after getting powerful physical evidence in cognitive research. This helps us understand the mechanisms in the brain of a human when dealing with different kinds of imperfections in information. The advent of methods like the Bayesian neural network and the neuro-fuzzy system is another reason for taking this approach. The start and progress in the research on the relationship between situation theory and neural networks will lead us to more powerful kinds of cognitive architectures. We must remember that deep learning algorithms have an important role in implementing some kinds of classification in this cognitive architecture.